\documentclass[11pt]{article}
\usepackage[a4paper, margin = 0.8in]{geometry}
\usepackage{graphicx}
\usepackage[dutch, english]{babel}
\usepackage{tcolorbox}
\usepackage{xcolor}
\usepackage{sectsty}

\usepackage{placeins}
\usepackage{amsmath, bm}
\usepackage{amssymb}
\usepackage{mathtools}
\usepackage{xurl}
\usepackage{textcomp}

\usepackage{amsthm}
\usepackage[sectionbib,square,numbers]{natbib}

\usepackage[font=footnotesize]{caption}
\usepackage[font=footnotesize]{subcaption}

\definecolor{MyBlue}{rgb}{0.2,0.2,0.6}
\sectionfont{\color{MyBlue}}
\subsectionfont{\color{MyBlue}}

\title{The complex relationship between anti-immigrant sentiment and exposure in the Netherlands}
\author{Benedikt Meylahn\textsuperscript{1,2,*}, Tommaso Giommoni\textsuperscript{1,3}, Mike Lees\textsuperscript{1,4}, Alessandro Nai\textsuperscript{5}}

\begin{document}
\maketitle
\begin{abstract}
\noindent
We study the relationship between exposure to immigrants and right-wing votes used as a proxy for anti-immigrant sentiment in the Netherlands. Exposure to immigrants is measured using a novel indicator, adapted from the residential segregation literature, applied to grid data available for the entire country at the aggregate level. Using the proxy of election results, we are able to study the entire voting population instead of a limited subset of panel respondents. In our cross-sectional study of the 2023 national election, we find that municipalities and neighbourhoods with greater exposure have less anti-immigrant sentiment. However, looking at the increase in right-wing vote share from the 2021 to the 2023 national elections, we find evidence of an ``overexposure effect'', as this relationship reverses, and the areas that saw significant increases in migrants subsequently experienced an increase in right-wing votes. Furthermore, we see that at high levels of exposure, wealth (and inequality) plays an important role in the voting dynamics, indicating that relative deprivation theory may play a role.

\vspace{1em}
{\sc 
\noindent Affiliations}: \\
\textsuperscript{1}Dutch Institute for Emergent Phenomena, University of Amsterdam.\\
\textsuperscript{2}Institute of Theoretical Physics, Institute of Physics, University of Amsterdam.\\
\textsuperscript{3}Quantitative Economics, Faculty of Economics, University of Amsterdam.\\
\textsuperscript{4}Computational Science Lab, Informatics Institute, University of Amsterdam.\\
\textsuperscript{5}Amsterdam School of Communication Research (ASCoR), University
of Amsterdam.\\
\textsuperscript{*}Corresponding author contact: b.v.meylahn[at]uva.nl

\vspace{1em}
{\sc 
\noindent Keywords}: Contact theory; Threat theory; Right-wing voting; Dutch elections%\\
%Word count: 9457
\end{abstract}

\section*{Significance statement}
Against the backdrop of worryingly fraught social cohesion and entrenched segregation into politically homogeneous factions, the relationship between exposure to immigrants and anti-immigrant sentiments is under debate. We study this relationship using the full voting population of the Netherlands as a robust proxy for right-wing voters' anti-immigrant sentiment. We show that the relationship follows the direction expected under contact theory on a national scale. However, upon closer inspection of individual regions, we show that the relationship may turn around, indicating that contact theory may tell part of the story but not the complete story. Furthermore, we illustrate the use of an exposure measure that takes distances into account in the context of minority group exposure and native group sentiment.
%\newpage

\section{Introduction}
Support for far-right, populist, anti-immigration political parties is on the rise in Europe. In the 2023 Dutch general election, the Party for Freedom (PVV) received 23\% of the vote, besting the runners-up by more than 7\% of the votes.\footnote{In the subsequent 2025 election they did not maintain this margin, yet their support stayed at 16.66\% losing only to the winners D66 at 16.94\%.} The surprise of this result is exacerbated when one considers that there was a decline in anti-immigration sentiment from 2017 to 2020 in the Netherlands~\cite{MUIS2022}.

Whether increased exposure to immigrants leads to an increase or decrease in anti-immigration sentiment is a question at the heart of an ever growing body of literature (see for instance~\cite{hainmueller2014public} for a review). Two theories lie at the core of this research, with empirical support for both: Group contact theory~\cite{allport1954nature} posits that exposure between groups has a positive effect on social relations between those groups. In contrast to this, threat theory~\cite{Putnam2007} suggests that a greater level of diversity leads to a breakdown of trust and cooperation. Two meta-reviews seem to indicate that there is not yet a consensus as to how these two theories can be reconciled. On the one hand, Pettigrew and Tropp~\cite{pettigrew2006meta} suggest that empirical support is abundant for group contact theory, while Kaufmann and Goodwin~\cite{Kaufmann2018} some years later indicate that there is more support for threat. They also go further to say that this may depend on the scale at which the analysis takes place. This is echoed by the findings of Lebow \textit{et al.}~\cite{Lebow2024} who show that while South American countries receiving the greater share of Venuzuelean immigrants showed a bigger increase in far-right support, within those countries, it was not the regions receiving the immigrants where this support surfaced.

\subsection{Related literature}
A full review of the literature is beyond the scope of this article. We focus on results and work that study the context of our own study: The Netherlands. Arzheimer \textit{et al.}~\cite{Arzheimer2024} studied similar dynamics in France, Germany, Great Britain, and the Netherlands, using survey data. Studies have also looked at Sweden (see~\cite{rydgren2013contextual}), South America (see~\cite{Lebow2024}), the UK and Japan (see~\cite{Igarashi2021}) and the islands in the Aegean Sea (see~\cite{hangartner2019exposure}). For a meta-analysis see~\cite{Kaufmann2018}, and for a review see~\cite{hainmueller2014public}.

Our work continues the investigation in the Netherlands, where Achard~\textit{et al.}~\cite{Achard2024} have shown a causal link between exposure to refugees (studying neighbourhoods with asylum centres and those without as control) and decreased anti-immigrant sentiments over the period 2011--2016. Similarly, Kazmina~\textit{et al.}~\cite{kazmina2024} study the effect of exposure in a population scale opportunity network on anti-immigration sentiment. They too find that increased exposure is related to more positive views on immigration, though they also find support for the notion that there exists a turning point after which more exposure increases anti-immigration sentiment. Moreover, van~Heerden and Ruedin~\cite{vanHeerden2019} have found a relationship between the change in the proportion of immigrants in Dutch neighbourhoods and anti-immigrant sentiment that corresponds to what would be expected for group contact theory.

In contrast to this, Gravelle \textit{et al.}~\cite{Gravelle2021} find evidence for group threat theory in the Netherlands in that the presence of mosques \textit{with} minarets is related to greater right-wing voting behaviour, while mosques without minarets had no such effect. This is echoed by later work with similar results in Switzerland~\cite{Valli2026}. 

In a meta-analysis Kaufmann and Goodwin~\cite{Kaufmann2018} observe that for many contexts there is more support in the literature for threat theory than for contact theory. They go on to relate the findings to the scale at which the study takes place. This, however, does not explain the disparity in results for the Dutch context. On the one hand we have causal contact theory links laid by Achard~\textit{et al.}~\cite{Achard2024}. On the other hand, Gravelle~\textit{et al.}~\cite{Gravelle2021} find evidence that increased exposure to visible mosques is related to greater right-wing, anti-immigrant voting. In between the two, Kazmina \textit{et al.}~\cite{kazmina2024} suggest there is contact up to a certain point of increased exposure, and threat thereafter. Finally, we have the results of the 2023 national election in the Netherlands, which indicate that for a decently sized portion of individuals, the feelings towards immigrants have not warmed over the period 2021-2023. 

Most of these studies are at the individual level, using the LISS panel (Longitudinal Internet studies for the Social Sciences) with roughly 5000 responses yearly. Our own study is on a regional scale. We leverage the strong connection between anti-immigrant sentiment and voting behaviour to use election results as a proxy for anti-immigration sentiment. Thus, our study includes the entire voting population of the Dutch population during the 2023 election, which, as mentioned before, witnessed a surge of far-right voting for the far-right PVV. With our analysis, we try to understand the patterns in the upset election of 2023 and, where possible, reconcile various findings in the literature in the context of the Netherlands.

\subsection{Contribution}
In this paper, we apply a measure of exposure from the study of segregation to the question of the effect of exposure to immigrants on the anti-immigration views of a local population. This measure has been used to study the residential segregation of republicans and democrats in the United States~\cite{Brown2021}, and is well suited to distinguish between two regions with differing spatial patterns (say fully-mixed vs.\,completely segregated) but an equal ratio of native to immigrant population. This is an innovation in this field, where studies typically use the share of the immigrant population as a measure of exposure. In well mixed neighbourhoods, the exposure measure we use and the share will not differ significantly, though for highly segregated regions, these two can be far apart (see the illustration in~\cite{Brown2021}).

Many studies focus on individuals and their responses to surveys regarding immigrants (for example~\cite{vanHeerden2019, Achard2024, kazmina2024}). This has its advantages; however, two disadvantages are that individuals may respond according to how they would like to be perceived rather than what they truly believe. Secondly, it restricts the reach of the study to those individuals responding to the survey. By using election data, we bypass these issues. Studies that use election data, however, are often limited by how they measure exposure. For instance, in Gravelle \textit{et al.}~\cite{Gravelle2021}, the inventive measure of mosques (with and without minarets) was used as a measure for exposure. While this certainly facilitates the study of specific dynamics related to, for instance, anti-Islam dynamics, it limits the scope of the region under study by the availability of data and the presence of mosques. Our granular (500m by 500m grid) data, combined with the innovative distance weighted exposure metric, overcomes these challenges, allowing us to study essentially the whole voting population of the Netherlands.

We are able to reconcile many of the findings in the literature related to group and contact theory in the Netherlands. This is possible because of the large scope we are able to study. We find that the relationship between exposure and right-wing voting typically follows the patterns of contact theory, with important caveats. The relationship between exposure and the \textit{increase} in right-wing voting follows a similar contact pattern to a point. Here we observe a turning point, a level of exposure after which more exposure leads to a greater increase in right-wing voting. Regions in which contact breaks down and right wing voting increases are typically rural, poor, and possibly in the presence of highly visible reminders of immigrants, such as improvised centres set up to deal with the influx of refugees and asylum seekers.

\section{Methods}

% The income data will be published later this year (April is the expectation).

\subsection{Exposure}\label{sec:exposuremethod}
\begin{figure}
    \centering
    \begin{subfigure}{0.45\textwidth}
        \includegraphics[width=\linewidth]{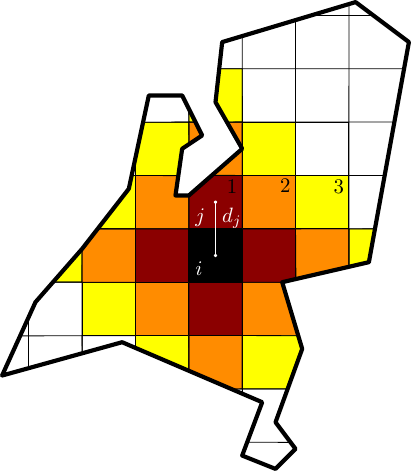}
    \caption{Exposure method illustrated}
    \label{fig:Expos_grid}
    \end{subfigure}%
    \begin{subfigure}{0.55\textwidth}
        \includegraphics[width=0.9\linewidth]{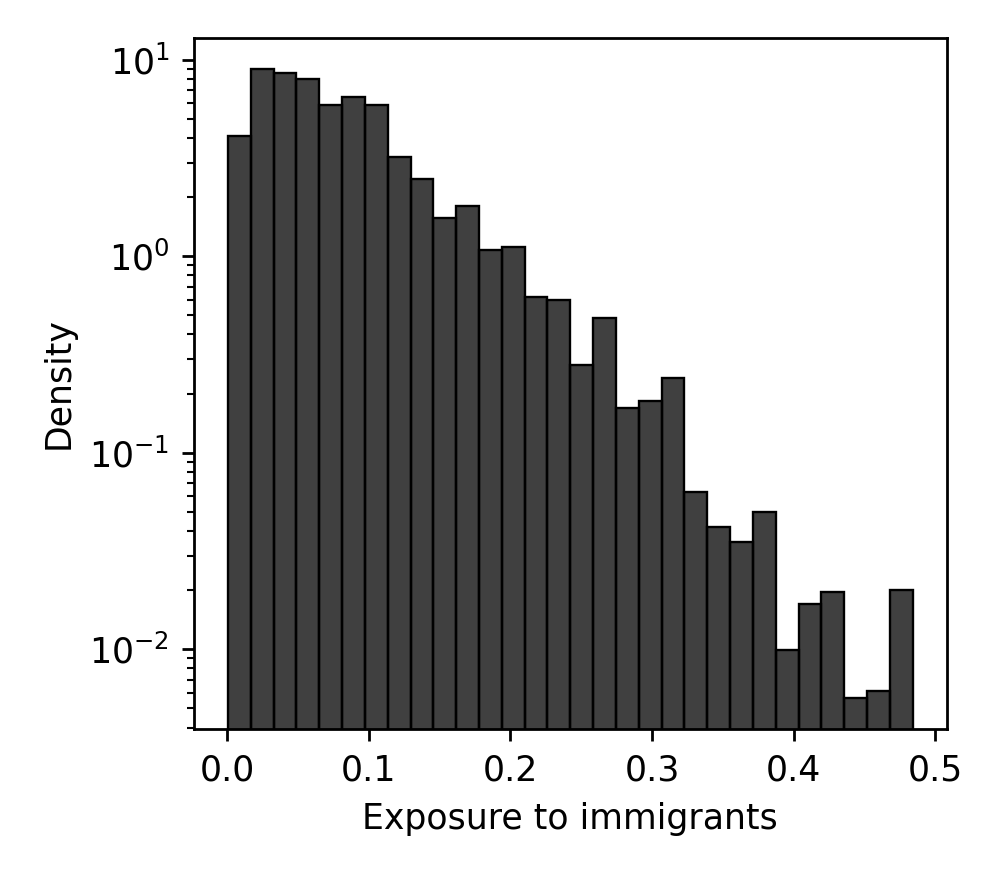}
    \caption{Mean exposure at neighbourhood level}\label{fig:histogram_exposure}
    \end{subfigure}
    \caption{(a) Illustration of exposure method on a grid. The cell for which exposure is calculated is labelled $i$ and coloured in black. The layers are labelled $1,2,3$ and coloured dark red, orange, and yellow respectively. The calculation starts with the contribution of the centre cell itself. (b) Histogram of mean exposure per native at neighbourhood scale. Note the y-axis is on log-scale. The histogram resembles an exponential distribution, though no formal checks are performed.}
\end{figure}

The way we calculate exposure to immigrants (and isolation of the Dutch population) is adapted from Brown and Enos~\cite{Brown2021} to suit the grid-data we have at our disposal. The exposure $E_i$ of Dutch people in cell $i$ to immigrants (defined as `born outside of the Netherlands, with a heritage outside of Europe'):
\begin{equation}
\label{eq:Expos_cell}
  E_i=\frac{\sum_{j\in N(i)}^{k}\frac{P^O_{j}}{d_j+c}}{\sum_{j\in N(i)}^{k}\frac{P_j}{d_j+c}}.
\end{equation}
Here:
\begin{itemize}
    \item $P^O_j$ is the number of residents of cell $j$ of the immigrant population,
    \item while $P_j$ is the total population of cell $j$, 
    \item $d_j$ is the (Euclidean) distance between cells $i$ and $j$, and
    \item $c$ is a constant tuning sensitivity to distance.
    \end{itemize}
The sum is taken over consecutive (Manhattan geometry) layers of neighbouring cells (starting with the cell in question) until a total population (not immigrant count) of $k$ has been reached. In Figure~\ref{fig:Expos_grid} we illustrate three of these layers from an illustrative cell $i$. 

For a neighbourhood or municipality (the units of study), we aggregate the exposure weighted by the number of Dutch persons in each cell.
\begin{equation}
    E_L = \frac{\sum_{l\in L}P^{NL}_l X_l}{\sum_{l\in L}P^{NL}_l},
\end{equation}
The sum is taken over all cells that are within the region in question. We then study the average or mean exposure experienced by a Dutch person in the region $L$.

\begin{figure}[hb!]
    \centering
       \begin{subfigure}{0.5\textwidth}
    \includegraphics[width=0.9\linewidth]{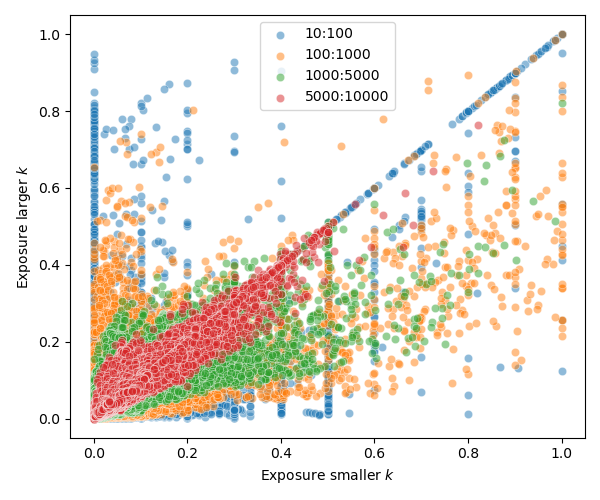}
    \caption{Grid}
    \label{fig:grid_exp_k_ratio}
    \end{subfigure}%
    \begin{subfigure}{0.5\textwidth}
    \includegraphics[width=0.9\linewidth]{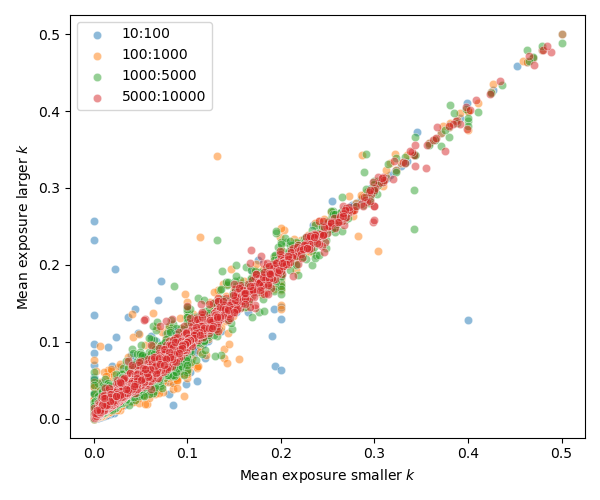}
    \caption{Neighbourhood}
    \label{fig:wijk_exp_k_ratio}
    \end{subfigure}%
    \caption{Comparison between exposure as calculated for different values of $k\in\{10,100, 1\,000, 5\,000, 10\,000\}$. The smaller of the $k$ values in the comparison is always on the $x$-axis.}
    \label{fig:exposure_k_robust}
\end{figure}

In Brown and Enos~\cite{Brown2021} these sums are taken over individuals and stop when exactly $k$ individuals have been reached. Due to the aggregated nature of grid data, we stop in the layer in which $k$ individuals are reached, but finish the entire layer to avoid errors related to how cells are ordered. This poses only a small threat, as the measure is quite robust to different values of $k$. This is because as more individuals are added, the distance to them increases, and so they provide a diminishing contribution to the total exposure. This is demonstrated in Figure~\ref{fig:exposure_k_robust}, where for both the grid level and the aggregated neighbourhood level, one can see that as $k$ increases, the scatter shifts closer to the $y=x$ line. In particular for the neighbourhood level this happens swiftly, indicating that the points far away from $y=x$ on the grid level are low density grid cells, which count less to the overall exposure in the population weighted aggregate level.

\subsection{Description of control variables}

\begin{itemize}
    \item \texttt{WOZ value}, the population weighted mean of the average `WOZ' value of homes in the grid cell. WOZ value of a home is its assessed Value under the Valuation of Immovable Property Act. We include this to account for differences in wealth that relate to property values.
    \item $\texttt{S}_{x,y}$ accounts for the age demographics of the regions and is the share of residents in the categories reported (\texttt{x, y})~$\in\{[15, 25), [25, 45), [45, 65)\}$. The share is calculated by dividing the count by the number of residents aged $\geq 15$ to focus on the voting population.
    \item \texttt{Address density}, the population weighted mean address density (`\foreignlanguage{dutch}{\textit{omgevingsaddressendichtheid}}'). This serves as a proxy for urbanisation with a high density found in cities and a low density in rural areas. This variable is reported as the mean number of addresses within 1 kilometer radius from addresses within the grid cell. 
\end{itemize}

Note that for the regression models we logarithmically scale and centre both the \texttt{address density} and the \texttt{WOZ value}. These are positive by definition and are both skewed, thus best log-transformed and then centred. This transformation is illustrated in Appendix~\ref{app:vars}, and gets applied after the population weighting.

The control variable for age we create is made to represent the share of individuals that are of voting age. We are unable to control for turnout by age and use the data available, the shares are therefore calculated as follows:
\begin{equation}
\begin{aligned}
    \texttt{S}_{15,25} &= \frac{N_{15,25}}{N_{15,25}+N_{25,45}+N_{45,65}+N_{65+}}\\
    \texttt{S}_{25,45} &= \frac{N_{25,45}}{N_{15,25}+N_{25,45}+N_{45,65}+N_{65+}}\\
    \texttt{S}_{45,65} &= \frac{N_{45,65}}{N_{15,25}+N_{25,45}+N_{45,65}+N_{65+}},
\end{aligned}
\end{equation}
Note that $S_{65+}$ is omitted as reference to avoid multi-collinearity. Furthermore the variables $N_{x_1,x_2}$ are the CBS variables reporting the number of individuals residing in that cell of age $\in [x_1,x_2)$, while $N_{65+}$ is the variable reporting the number of people aged 65 and older. 

\subsection{Population weighted control variables}
Similarly to how we take the weighted mean of exposure, where the weight is the number of Dutch people experiencing the exposure, we take population weighted means of the control variables of WOZ home value and the Address density.

The weighted regional (be it municipality or neighbourhood) population address density (or WOZ home value) is calculated as:
\begin{equation}
    \texttt{Address density}_L = \frac{\sum_{l\in L}P_l \cdot \texttt{Address density}_l}{\sum_{l\in L}P_l},
\end{equation}
where $P_l$ is the population in the cell $l$ and the sum is taken over all cells $l$ in the region $L$. The same is true for the WOZ value with the appropriate replacements. 

\subsection{Data, units of analysis and visualisations}
The units of analysis we study are the municipalities and the neighbourhoods defined by the Statistics Netherlands (CBS) `\textit{gebiedsindelingen}'~\cite{cbs_gebiedsindelingen_2025} which are simplified versions of the actual boundaries for cartographical purposes. We populate these regions with the 2023 (version 2) 500m-by-500m grid data~\cite{cbs_vierkanten} aggregated along the aforementioned neighbourhood and municipality boundaries. All computations and aggregations are our own and not attributed to Statistics Netherlands (CBS).

In order to determine which parties are to be considered `right-wing, anti-immigrant' we study the Longitudinal Internet studies for the Social Sciences (LISS) panel~\cite{Scherpenzeel2011} section on politics and values (waves 2016--2025) this is outlined in Appendix~\ref{app:liss_check}. The data~\cite{liss_politics_values_2022} are managed by CentERdata (Tilburg University, the Netherlands) and are available at \url{www.dataarchive.lissdata.nl}.

When we account for regional fixed effects, we use the division of the Netherlands into COROP regions. These are used by Statistics Netherlands and are a NUTS 3 designation. In addition, we sometimes use the 12 Provinces of the Netherlands to compare the results using Provincial, COROP and no fixed effects.

\section{Results}
The relationship between exposure to minority group immigrants and the proxy for anti-immigrant sentiment (right-wing or PVV voting) shows a national trend, as is expected by contact theory. However, when drilling down into regions of similar urbanicity, this relationship partially breaks down. For the purpose of this paper, we define minority group immigrants as persons born outside of the Netherlands with a non-European heritage. The number of residents in this group is reported in the Statistics Netherlands (CBS) grid data and is chosen as the most representative of the group one would ideally study — those who would be readily perceived as foreign by the Dutch populace.

\subsection{Exposure to immigrants in the Netherlands}

If immigrants were evenly distributed throughout the Netherlands, the mean exposure felt by natives would be uniformly distributed. In Figure~\ref{fig:histogram_exposure}, plotting the histogram of mean exposure felt by Dutch people per neighbourhood, we see that this is far from the case. Instead, there are many neighbourhoods with low exposure and a few with much higher mean exposure. 

\begin{figure}
    \centering
    \includegraphics[width=0.98\linewidth]{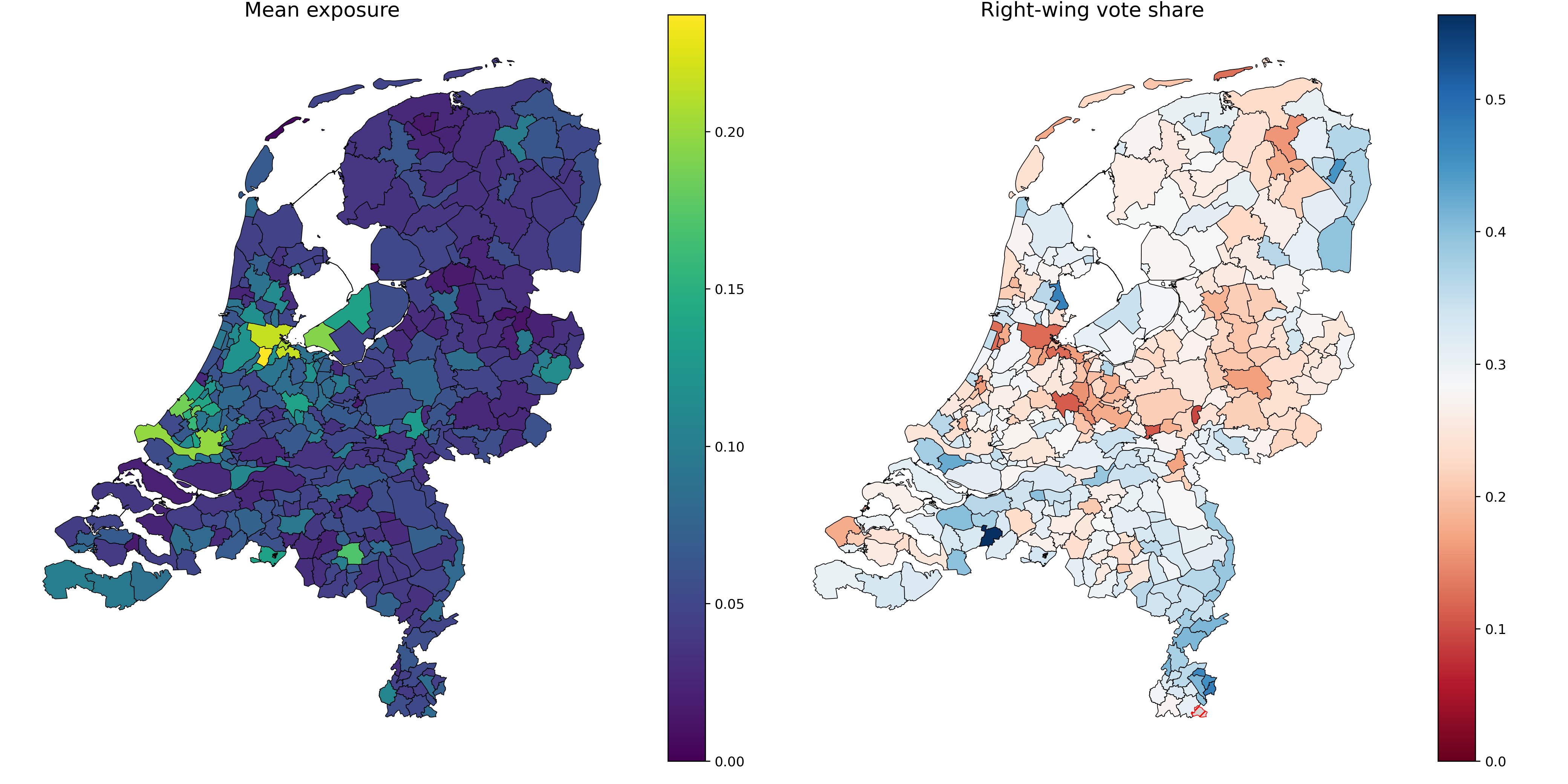}
    \caption{The exposure on average by a Dutch person per municipality in 2023 and the  right-wing vote share per municipality in 2023. Grey regions with a red border and hatching indicate a lack of data.}
    \label{fig:Expos_RW_Municipality}
\end{figure}

Looking at the spatial distribution of the exposure across municipalities in the left panel of Figure~\ref{fig:Expos_RW_Municipality}, we see again that exposure is far from uniform across the country. The most exposed municipalities have average exposure upwards of 0.2, while the least exposed are not far from zero exposure. We see the expected concentration in and around the large cities (Amsterdam, Rotterdam, The Hague), as well as around Eindhoven. Perhaps surprisingly, we see regions of relatively high exposure along the eastern and southern borders of the Netherlands.

\begin{figure}[htb]
    \centering
    \begin{subfigure}{0.5\textwidth}
        \includegraphics[width=0.9\linewidth]{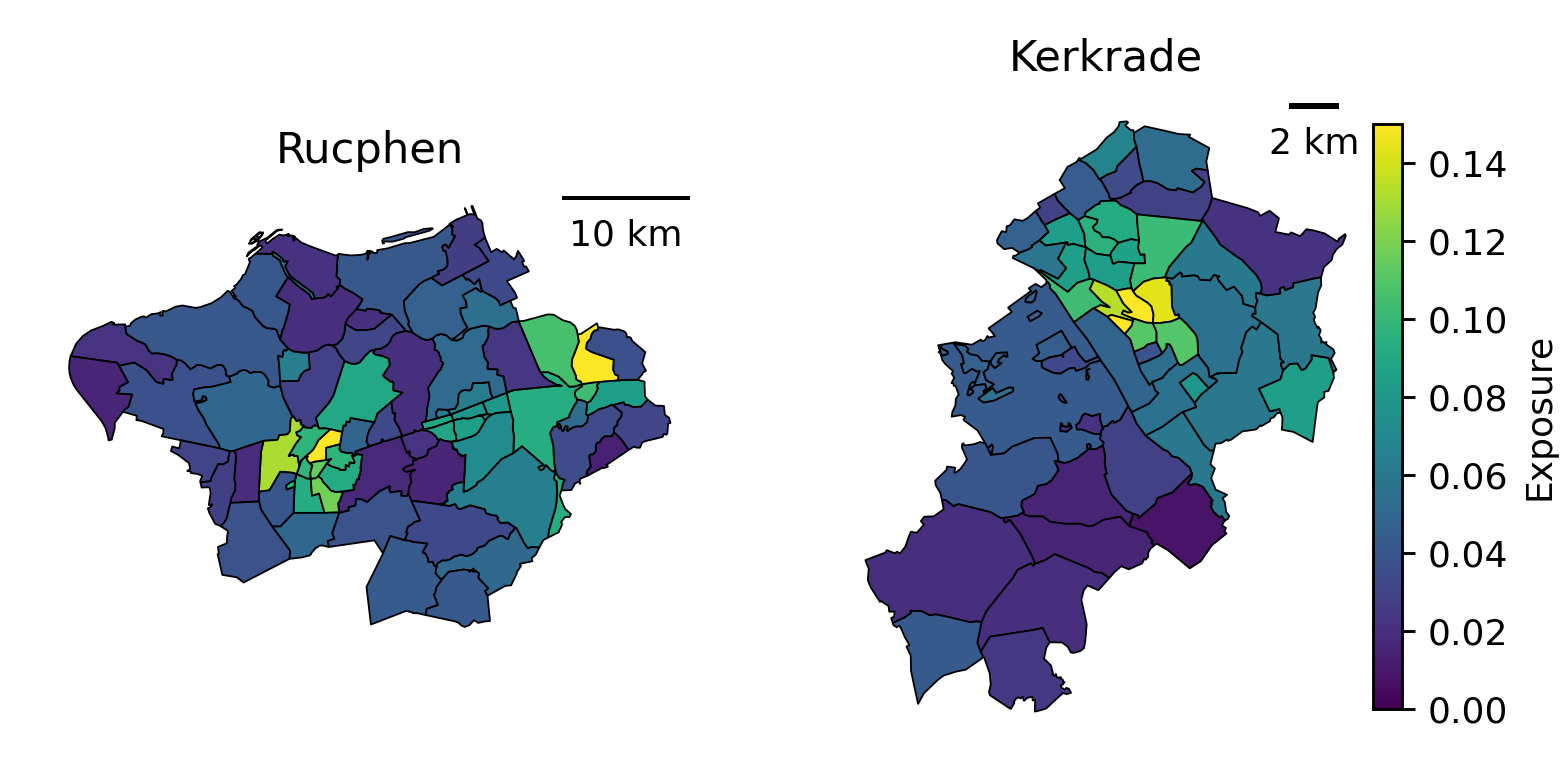}
    \caption{Exposure to immigrants.}
    \label{fig:RucKerk_Exp_neighbourhood}
    \end{subfigure}%
    \begin{subfigure}{0.5\textwidth}
        \includegraphics[width=0.9\linewidth]{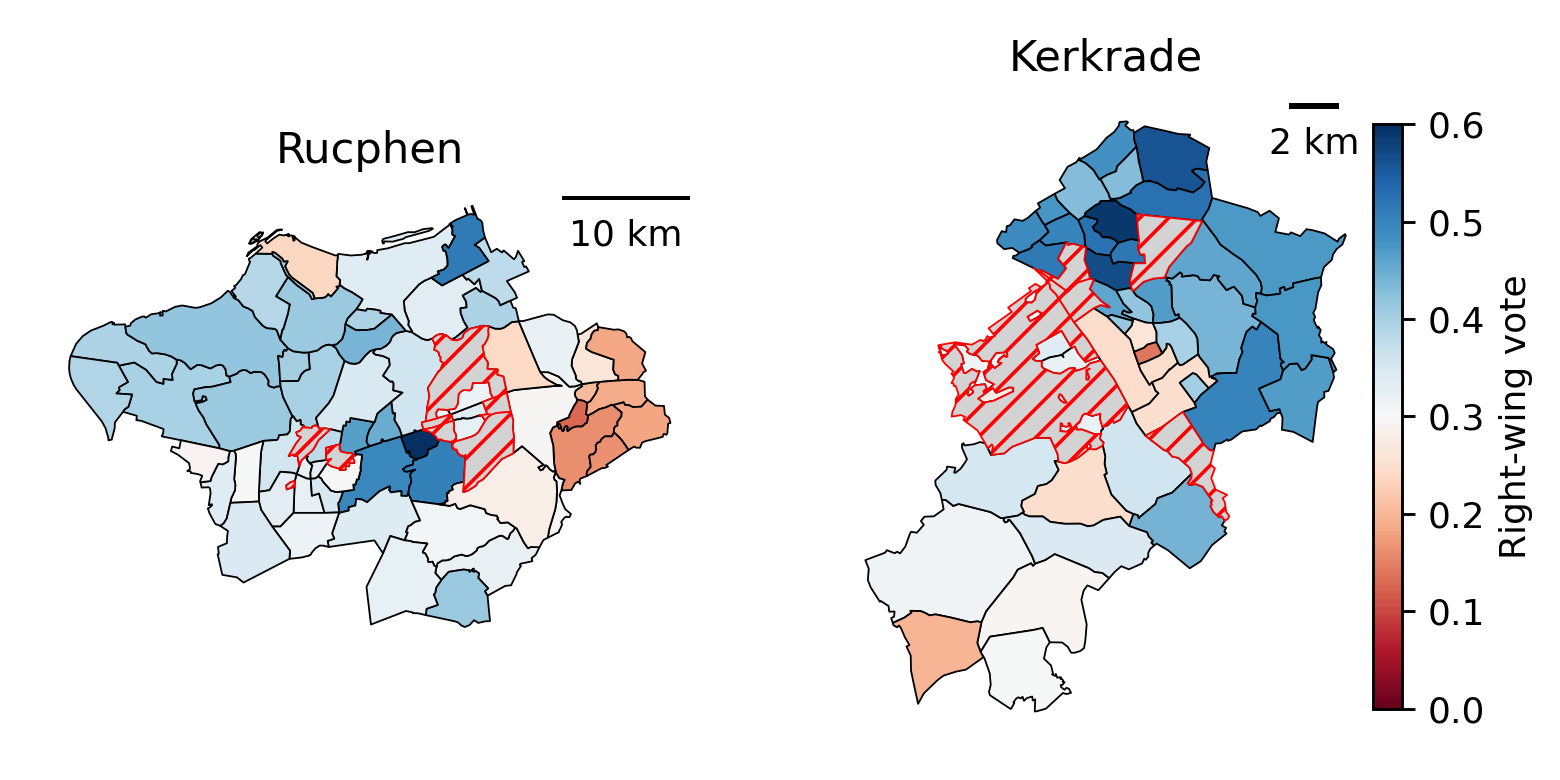}
   \caption{Right-wing voting.}
   \label{fig:RucKerk_right-wingvote}
    \end{subfigure}%
    \caption{Exposure and right-wing voting shares Rucphen and Kerkrade, selected based on the municipalities with the highest and second-highest support for right-wing, anti-immigrant parties. The sample includes all neighbourhoods within municipalities located within 5km of the borders. Right-wing voting in Rucphen (Sint-Willebrord) reaches 0.76; the colormap is truncated at 0.6 to make variation in Kerkrade visible on the same scale. Grey regions with a red border and hatching indicate a lack of data.}
    \label{fig:Exp_RW_zoom}
\end{figure}

At the municipal aggregate level (Figure~\ref{fig:Expos_RW_Municipality}), the share of right-wing votes and the exposure of immigrants demonstrate a pronounced negative correlation. However, this macro-spatial pattern dissipates at higher resolutions. When disaggregating the data to the neighbourhood level within the two municipalities that record the highest right-wing electoral support (Figure~\ref{fig:Exp_RW_zoom}), this inverse relationship is no longer discernible. Plotting localised exposure against right-wing vote share in these specific focal regions reveals a micro-spatial heterogeneity that is obscured by municipal-level averages.

\begin{figure}[htb]
    \centering
    \begin{subfigure}{0.5\textwidth}
        \includegraphics[width=0.9\linewidth]{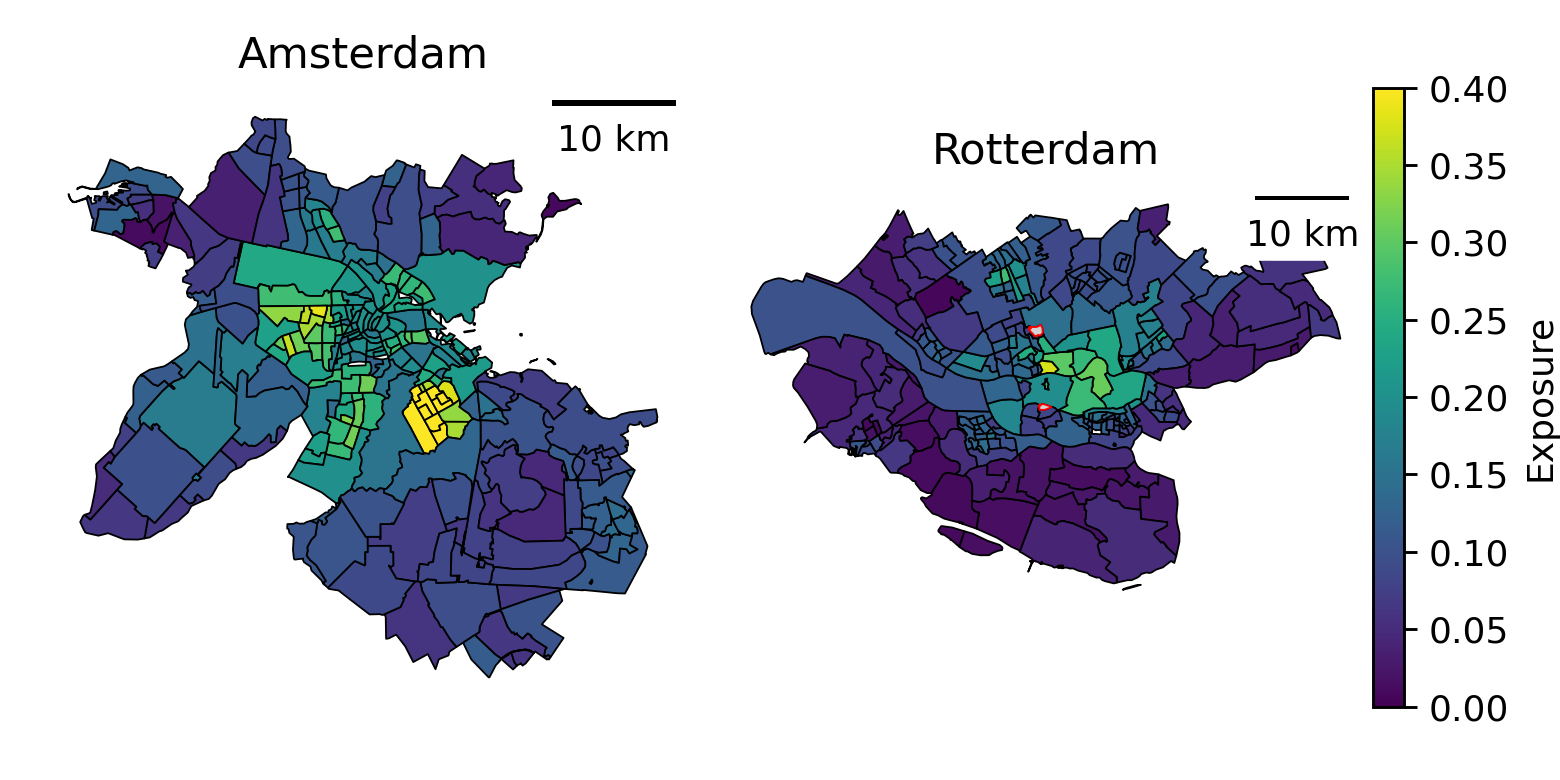}
    \caption{Exposure to immigrants.}
    \label{fig:AMS_ROT_Exp_neighbourhood}
    \end{subfigure}%
    \begin{subfigure}{0.5\textwidth}
        \includegraphics[width=0.9\linewidth]{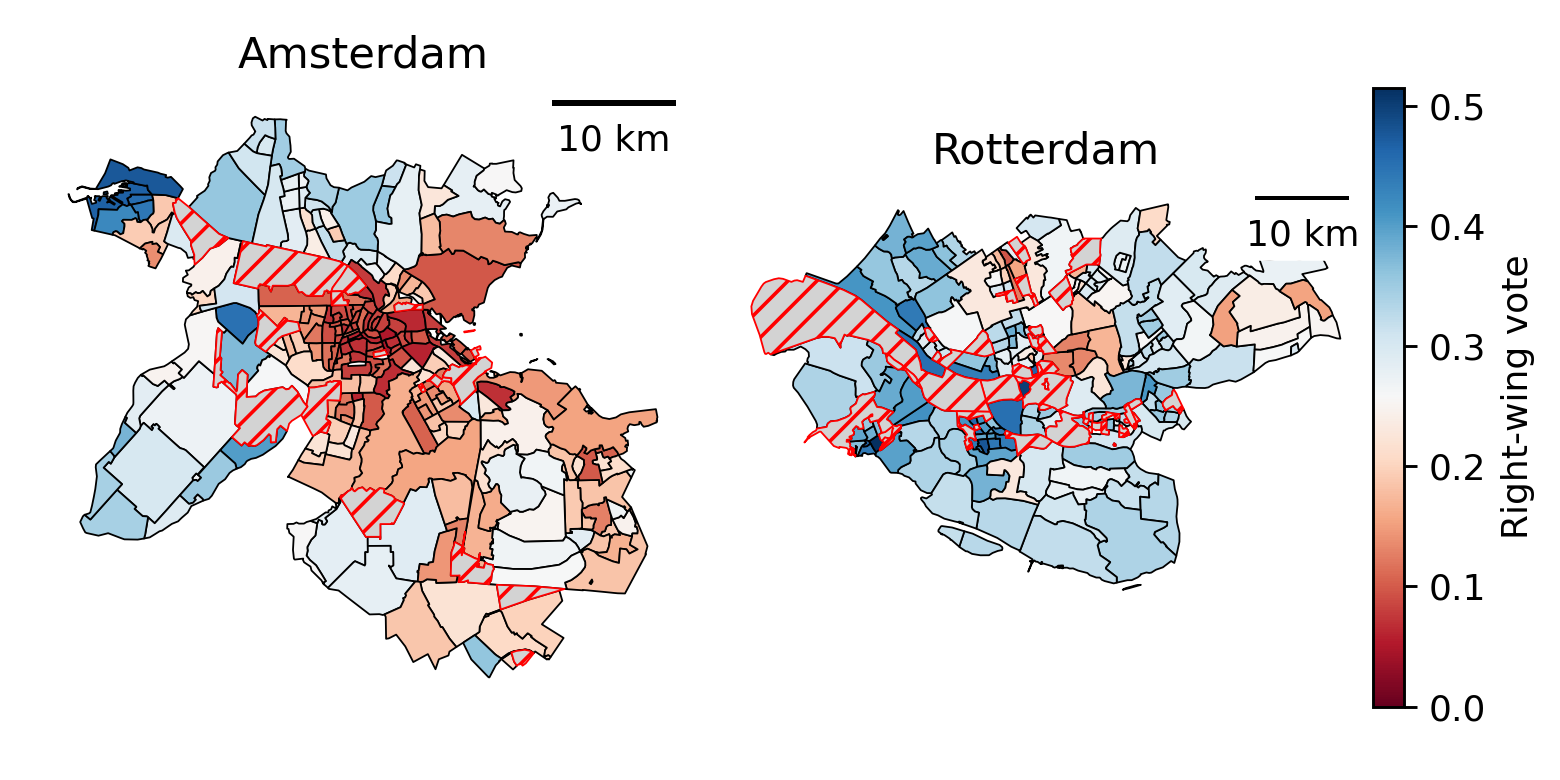}
   \caption{Right-wing voting.}
   \label{fig:AMS_ROT_right-wingvote}
    \end{subfigure}%
    \caption{Exposure and right-wing voting shares in Amsterdam and Rotterdam, selected based on the municipalities with the highest and second-highest exposure. The sample includes all neighbourhoods within municipalities located within 2km of the borders. Exposure in Amsterdam (H-buurt) reaches 0.5; the colormap is truncated at 0.4 to make variation in Rotterdam visible on the same scale. Grey regions with a red border and hatching indicate a lack of data.}
    \label{fig:Exp_RW_zoom_AmsRot}
\end{figure}

Finally, Figure~\ref{fig:Exp_RW_zoom_AmsRot} details the micro-spatial dynamics within the two municipal regions recording the highest aggregate immigrant exposure. Visual inspection of these areas does not readily corroborate the predictions of group threat theory. Rather, within these highly exposed urban centres and their immediate peripheries, neighbourhoods with the greatest localised exposure generally exhibit attenuated right-wing vote shares. In contrast, low-exposure neighbourhoods demonstrate significant electoral variance, capturing both high and low levels of right-wing support. To assess these heterogeneous results, the subsequent sections systematically quantify this relationship across varying spatial scales.

\subsection{Relationship between exposure and sentiment}
Following similar studies in the literature, we include various controls. Through these controls, we aim to account for the effects of wealth, age, urbanisation, and regional politics (by means of COROP regions as regional fixed effects). From the data, in some cases, we use proxies for these concepts. The control variables we include are described in the Methods section.

\begin{figure}[htb!]
    \centering
    \begin{subfigure}{0.5\textwidth}
        \includegraphics[width=0.95\linewidth]{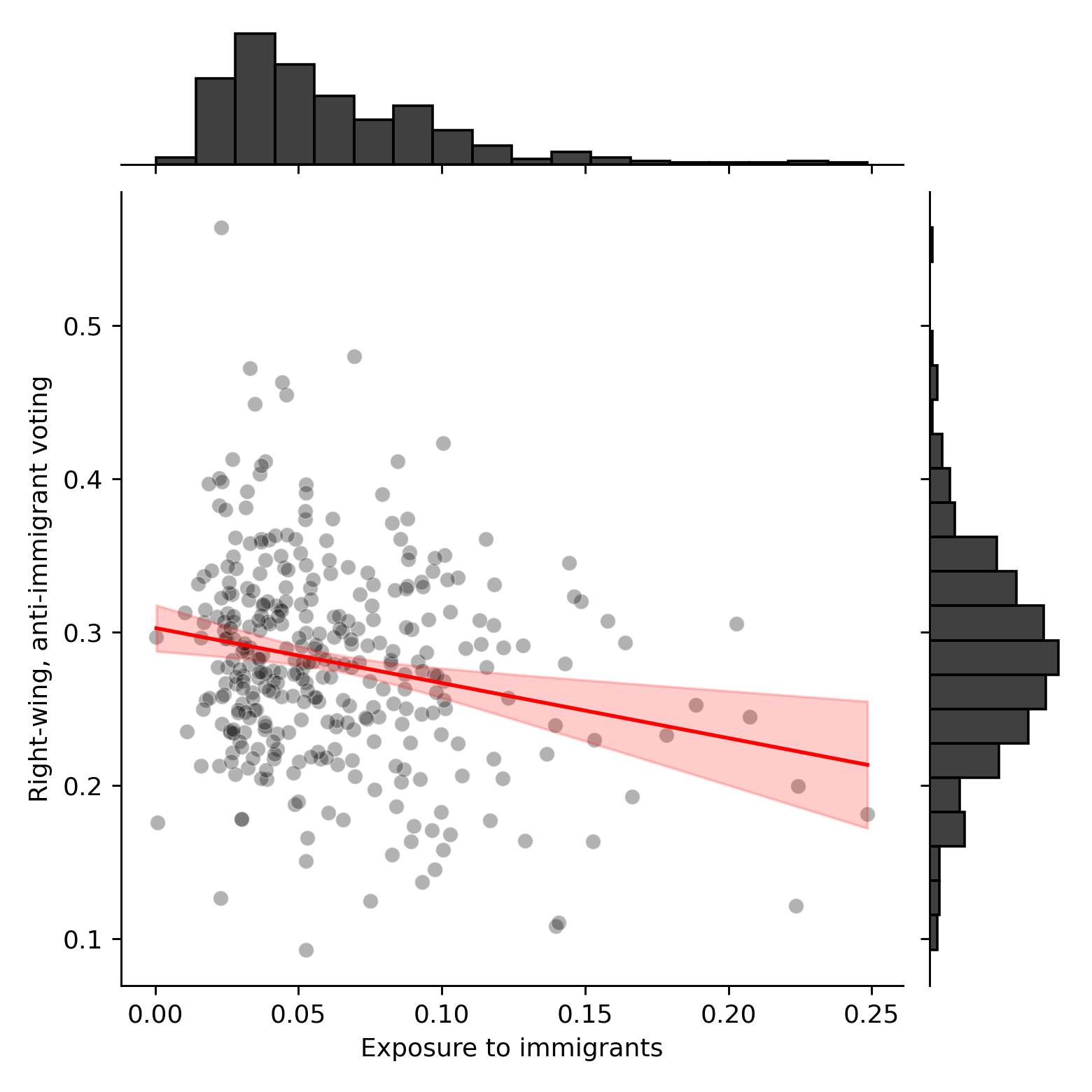}
    \caption{Municipality. slope: $-0.359, p = 0.002, N=341$.}
    \label{fig:exp_muni}
    \end{subfigure}%
    \begin{subfigure}{0.5\textwidth}
        \includegraphics[width=0.95\linewidth]{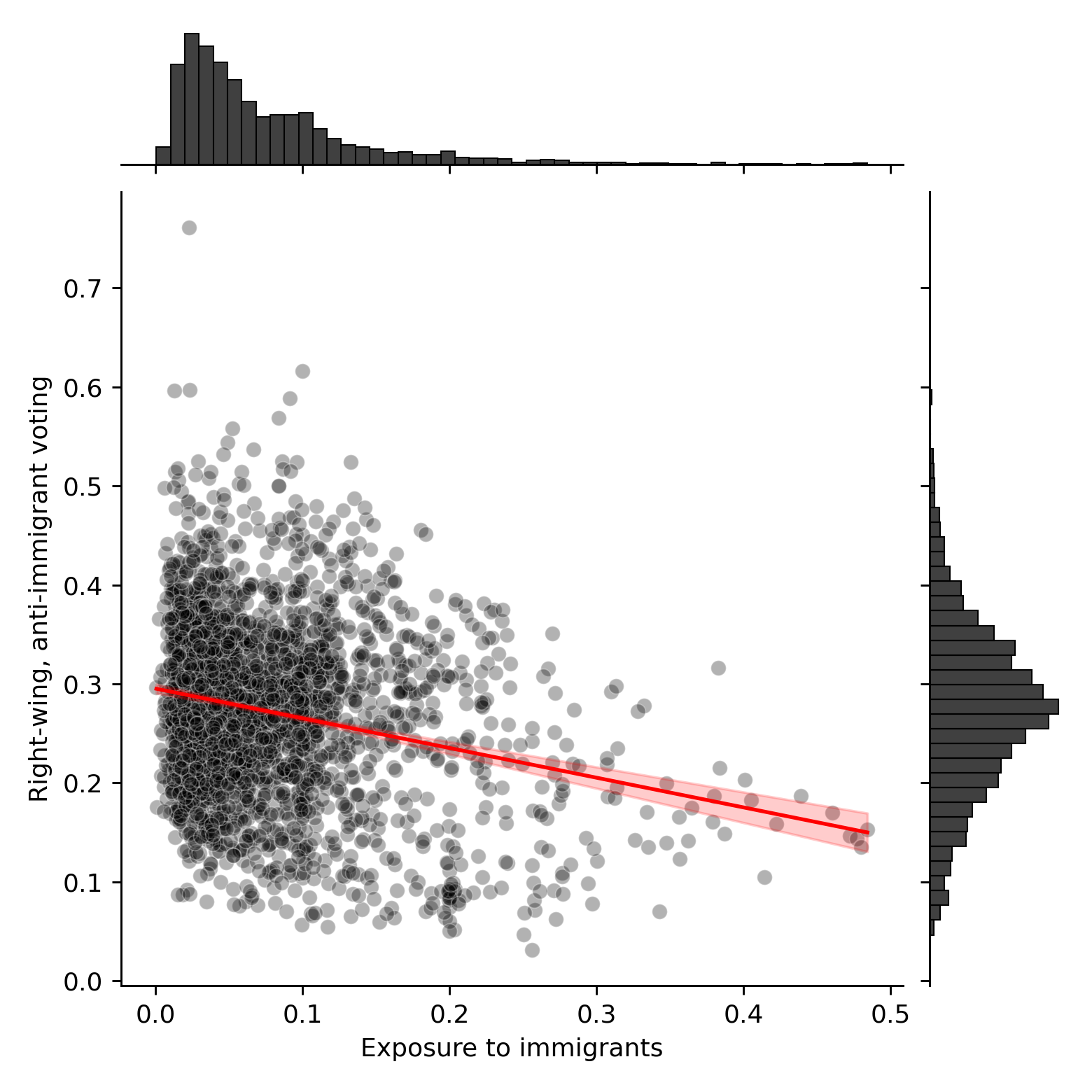}
   \caption{Neighbourhood. slope: $-0.300, p<0.001, N=2\,816$.}
   \label{fig:exp_wijk}
    \end{subfigure}%
    \caption{Exposure to immigrants and in relationship to right-wing, anti-immigrant voting per municipality and neighbourhood in models M1 and M2 in Table~\ref{tab:baseModelRegr} (in red is the marginal effect as per the regression model, assuming the controls take their mean value).}
    \label{fig:ExpIsoRW}
\end{figure}

In Figure~\ref{fig:ExpIsoRW} we show the relationship between exposure to immigrants and right-wing, anti-immigrant voting. In Figure~\ref{fig:exp_muni} unit of measurement is the municipality (Dutch: \textit{gemeente}), while in Figure~\ref{fig:exp_wijk} the unit is the neighbourhood (Dutch: \textit{wijk}). On the $y$-axis, we plot the share of the eligible votes that were awarded to the right-wing, anti-immigrant parties in voting venues located inside the neighbourhood in question. The $x$-axis represents the exposure experienced by the average Dutch person in the neighbourhood. The regression line corresponds to the marginal effect from the ordinary least squares regression, with controls held at their mean values.

We see that at a national scale the relationship predicted by contact theory is observed: Regions with more exposure to immigrants exhibit less right-wing, anti-immigrant voting.

\begin{table}[htb!]
\caption{Regression results on both the municipality and the neighbourhood scales (including insignificant coefficients). For all models ordinary least squares (OLS) regression coefficients are shown. The $k$ value used to compute exposure is $10\,000$. Standard errors in parentheses. FE: Fixed effects.}
\label{tab:baseModelRegr}
\begin{center}
\resizebox{\textwidth}{!}{
\begin{tabular}{l|ll|ll|ll}
\hline

                            & \multicolumn{2}{c}{No Fixed effects} & \multicolumn{2}{c}{Province FE} & \multicolumn{2}{c}{COROP FE}  \\
                            & Municipality & Neighbourhood & Municipality & Neighbourhood & Municipality & Neighbourhood  \\
                            & M1 & M2 & M3 & M4 & M5 & M6 \\

\hline
Exposure                    & -0.359***                    & -0.300***                & -0.287***         & -0.348***     & -0.496***        & -0.421***     \\
                            & (0.114)                      & (0.024)                  & (0.106)           & (0.024)       & (0.128)          & (0.030)       \\[4pt]
{Address density}    & 0.012**                      & 0.021***                 & -0.001            & 0.004*        & 0.000            & 0.004         \\
                            & (0.005)                      & (0.003)                  & (0.004)           & (0.002)       & (0.004)          & (0.003)       \\[4pt]

{WOZ value}          & -0.035***                    & -0.050***                & -0.048***         & -0.063***     & -0.057***        & -0.067***     \\
                            & (0.003)                      & (0.001)                  & (0.004)           & (0.002)       & (0.005)          & (0.002)       \\[4pt]
$\texttt{S}_{15,25}$        & -0.690***                    & -0.262***                & -0.108            & -0.166***     & -0.061           & -0.131***     \\
                            & (0.212)                      & (0.042)                  & (0.171)           & (0.039)       & (0.170)          & (0.038)       \\[4pt]
$\texttt{S}_{25,45}$        & 0.553***                     & 0.093***                 & 0.342***          & 0.049**       & 0.157            & 0.037         \\
                            & (0.165)                      & (0.028)                  & (0.130)           & (0.025)       & (0.133)          & (0.025)       \\[4pt]
$\texttt{S}_{45,65}$        & 1.239***                     & 0.550***                 & 1.186***          & 0.501***      & 0.957***         & 0.497***      \\
                            & (0.231)                      & (0.058)                  & (0.189)           & (0.055)       & (0.186)          & (0.056)       \\[10pt]
                            
R-squared                   & 0.425                        & 0.369                    & 0.618             & 0.530         & 0.726            & 0.596         \\
R-squared Adj.              & 0.415                        & 0.368                    & 0.598             & 0.527         & 0.684            & 0.589         \\
N                           & 341                          & 2816                     & 341               & 2816          & 341              & 2816          \\[4pt]
\hline
\multicolumn{3}{l}{*$p<.1$, **$p<.05$, ***$p<.01$}
\end{tabular}}
\end{center}
\end{table}

Table~\ref{tab:baseModelRegr} shows the coefficients of the multiple regression of model of the form:
\begin{equation}
    y_i  = \alpha E_i + \bm{\beta}\cdot\texttt{Controls}_i +\epsilon_i,
\end{equation}
where $y_i$ is the right-wing vote, $E_i$ is the mean exposure experienced by Dutch people in the region $i$, and the controls are the base controls we introduced above (counting for wealth, age, and urbanisation).

From the regression, it is clear that the global picture is in line with group contact theory: exposure is negatively associated with right-wing voting. This relationship is present in all the model specifications: On both municipal (M1, M3, and M5) and neighbourhood (M2, M4, M6) scales, without regional fixed effects taken into account (M1, M2), and with regional fixed effects taken into account (M3, M4 Province, and M5, M6 COROP region), this relationship holds.

Our primary empirical specification operates at the neighbourhood level and incorporates entity fixed effects to absorb unobserved spatial heterogeneity. This high-resolution approach captures localised exposure dynamics while using a substantial sample size (N = 2\,816 neighbourhoods) to ensure robust statistical power. Furthermore, Dutch administrative neighbourhoods represent substantively meaningful socio-spatial units that generally encompass the core of residents' routine daily activities and interpersonal encounters.\footnote{We note a structural exception in rural municipalities, where lower population densities often necessitate inter-neighbourhood travel for essential services such as secondary education or large-scale retail.} Therefore for the remainder of the paper, we compare new specifications to M6, which is at the neighbourhood scale. 

We have performed robustness checks of these findings against:
\begin{itemize}
    \item differing values of $k$ in the exposure measure (see Appendix~\ref{app:diff_k}),
    \item the inclusion of municipal voting behaviour in the last general election of 2021 as a control (see Appendix~\ref{app:historyMuni}),
    \item alternative definitions of right-wing, anti-immigrant voting (see Appendix~\ref{app:PVV}), and
    \item excluding the region with Ter Apel, the neighbourhood with the central arrival point for asylum seekers in the Netherlands (see Appendix~\ref{app:TerApel}).
\end{itemize}
The findings survive all of the above robustness checks.

\subsection{Interactions with address density and home value}
In the M6 we see that address density does not account for all of the difference in voting behaviour.\footnote{Actually, this is true for all the model specifications in Table~\ref{tab:baseModelRegr}.} In fact, the coefficients of address density are not even significant in all models of Table~\ref{tab:baseModelRegr} and their sign changes under different specifications. In the following analysis, we see that the interaction between address density and exposure is statistically significant. The magnitude and direction of the coefficient of exposure change as the setting becomes more or less urbanised. 

The home value (WOZ value) has a consistently negative and significant coefficient in all of the models presented so far. Regions with higher home values are associated with less right-wing, anti-immigrant voting behaviour. To further investigate, we examine the link between home value and exposure to identify whether the effect of exposure may depend on the economic situation in which it takes place.

\begin{table}[htb]
\caption{Interaction analysis for the effect of exposure with address density and home value. The analysis unit is the neighbourhood for all models. Note that here as before the Address density and the WOZ home value variables are first transformed to by the natural logarithm and then centred. All models in this table include the province to account for regional fixed effects.}
\label{tab:main_interactions}
\begin{center}
    \begin{tabular}{l|l|l|ll}
    \hline
 Interaction with                   & Address density  & Home value   & Both & Both  \\
  & M7 & M8 & M9 & M10\\
\hline
Exposure                                          & -0.046      & -0.413***   & -0.048         & 0.038          \\
                                                  & (0.040)     & (0.033)     & (0.041)        & (0.041)        \\[4pt]

{Address density}                          & 0.021***    & 0.003       & 0.021***       & 0.019***       \\
                                                   & (0.003)     & (0.003)     & (0.003)        & (0.003)        \\[4pt]
Exposure:{Address density}                 & -0.400***   &             & -0.400***      & -0.361***      \\
                                                  & (0.030)     &             & (0.030)        & (0.029)        \\ [4pt]
{WOZ value}                                & -0.067***   & -0.068***   & -0.067***      & -0.066***      \\
                                                  & (0.002)     & (0.003)     & (0.003)        & (0.003)        \\[4pt]
Exposure:{WOZ value}                       &             & 0.014       & -0.005         & 0.035*         \\
                                                  &             & (0.021)     & (0.021)        & (0.018)        \\[10pt]
% $\texttt{S}_{15,25}$                              & -0.098**    & -0.128***   & -0.099**       & -0.138***      \\

% $\texttt{S}_{25,45}$                              & 0.103***    & 0.035       & 0.104***       & 0.109***       \\

% $\texttt{S}_{45,65}$                              & 0.475***    & 0.496***    & 0.475***       & 0.483***       \\

Regional fixed effects                            &    COROP    &    COROP    &    COROP      &    Province       \\

R-squared                                         & 0.623       & 0.596       & 0.623          & 0.555          \\
R-squared Adj.                                    & 0.617       & 0.589       & 0.617          & 0.552          \\
N                                                 & 2816        & 2816        & 2816           & 2816           \\
\hline
\multicolumn{3}{l}{*$p<.1$, **$p<.05$, ***$p<.01$}
\end{tabular}
\end{center}

% *$p<.1$, **$p<.05$, ***$p<.01$
\end{table}
The coefficients of the interaction models (M7--M10) are presented in Table~\ref{tab:main_interactions}. We see that the baseline effect of increased address density (greater urbanisation) is more right wing voting. However, the interaction with exposure has a statistically significant negative coefficient. That is, exposure in regions of dense housing has a {\em stronger} negative relationship with right-voting. Increased exposure in low density regions, however, is associated with more right-wing voting. This is true for all model specifications that include an interaction with address density (M7, M9, and M10 in Table~\ref{tab:main_interactions}). 

% This is true across the model with only density interaction and the model with both interactions. 

The coefficients of interaction between home value (WOZ) and exposure are not consistent in terms of sign among M8, M9, and M10 and are not significant. We conclude that there is likely no significant interaction between home value and exposure to immigrants in right-wing voting. It therefore seems that wealth does not change the nature of the relationship between exposure and right-wing voting in the Netherlands. It should be noted, of course, that home value consistently has a significant negative correlation with right-wing voting. Richer regions have less anti-immigrant voting than their poorer counterparts. Our findings in this regard are in contrast to the review by Hainmueller and Hopkins~\cite{hainmueller2014public} who find that personal economic situations are not strongly related to immigration attitudes.

\begin{figure}[htb!]
    \centering
    \includegraphics[width=0.9\linewidth]{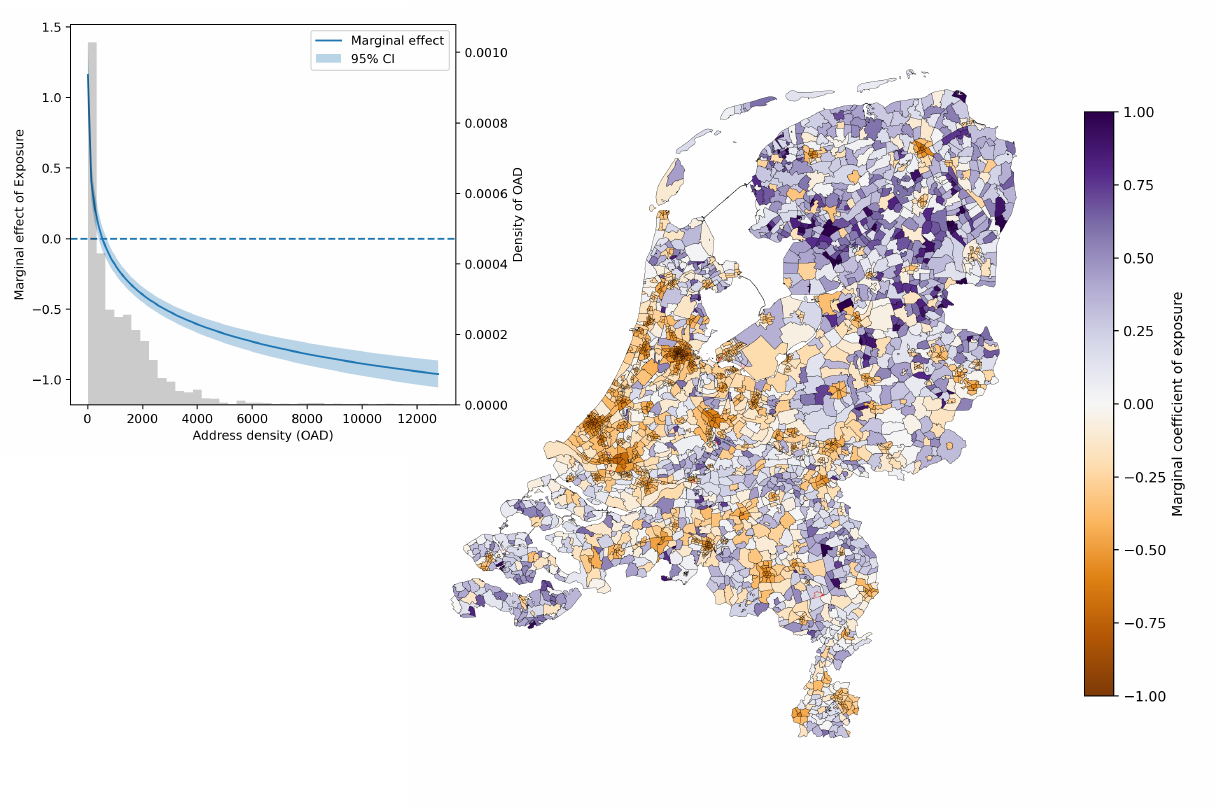}
    \caption{Marginal coefficient of exposure under the M7 in Table~\ref{tab:main_interactions} per neighbourhood. Inlay is the same plotted against address density with 95\% confidence intervals.}
    % Data: CBS Wijk- en Buurtkaart 2023. \textcopyright\,Kadaster / Centraal Bureau voor de Statistiek, 2024.}
    \label{fig:marginal_E_oad}
\end{figure}

Models that feature an interaction between address density and exposure have a marginal effect of exposure that changes sign for low density regions. We show the marginal effect in Figure~\ref{fig:marginal_E_oad} across the Netherlands per neighbourhood and against address density in the inlay. That is, we plot the value of:
\begin{equation}
    \frac{\partial y}{\partial E} = -0.046 -0.4 \cdot \texttt{Address density},
\end{equation}
taking the coefficients from the address density interaction model M7 in Table~\ref{tab:main_interactions}. The map shows that although the number of neighbourhoods that have a positive marginal coefficient of exposure is low, they make up large swathes of land between urban centres.

From the map, we see that there are indeed regions that have address density such that the marginal effect of exposure is an increase in right-wing voting. We also see this in the inlay, which has a histogram as a backdrop illustrating that both the positive and negative marginal coefficients of exposure are well represented in the data, and neither is an artefact of extrapolation.

\subsection{Relationship between exposure and the change in right-wing vote share from 2021 to 2023}
\begin{figure}[htb!]
    \centering
    \begin{subfigure}{0.5\textwidth}
        \includegraphics[width=0.95\linewidth]{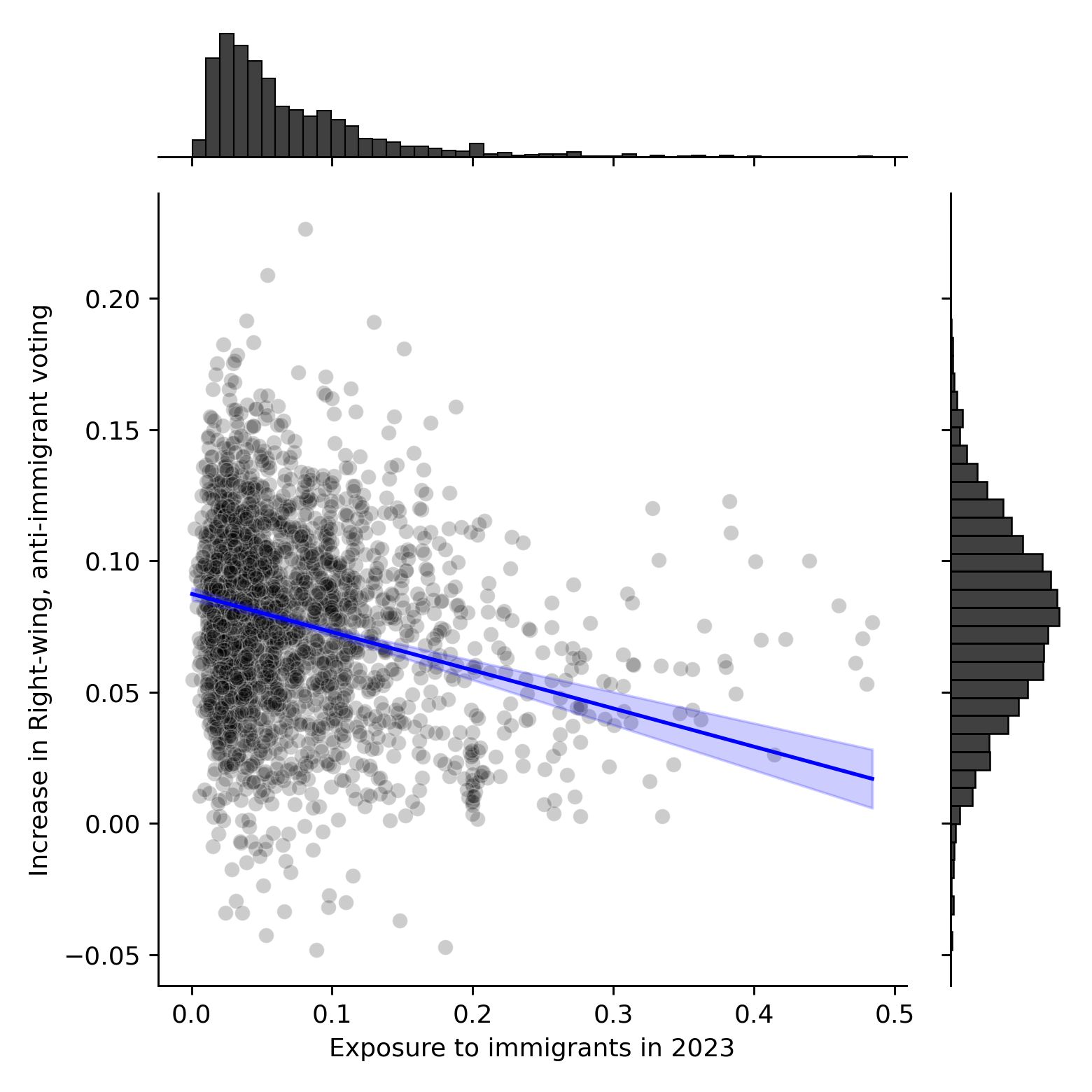}
    \caption{Linear (D1)}
    \label{fig:exp_diff_lin}
    \end{subfigure}%
    \begin{subfigure}{0.5\textwidth}
        \includegraphics[width=0.95\linewidth]{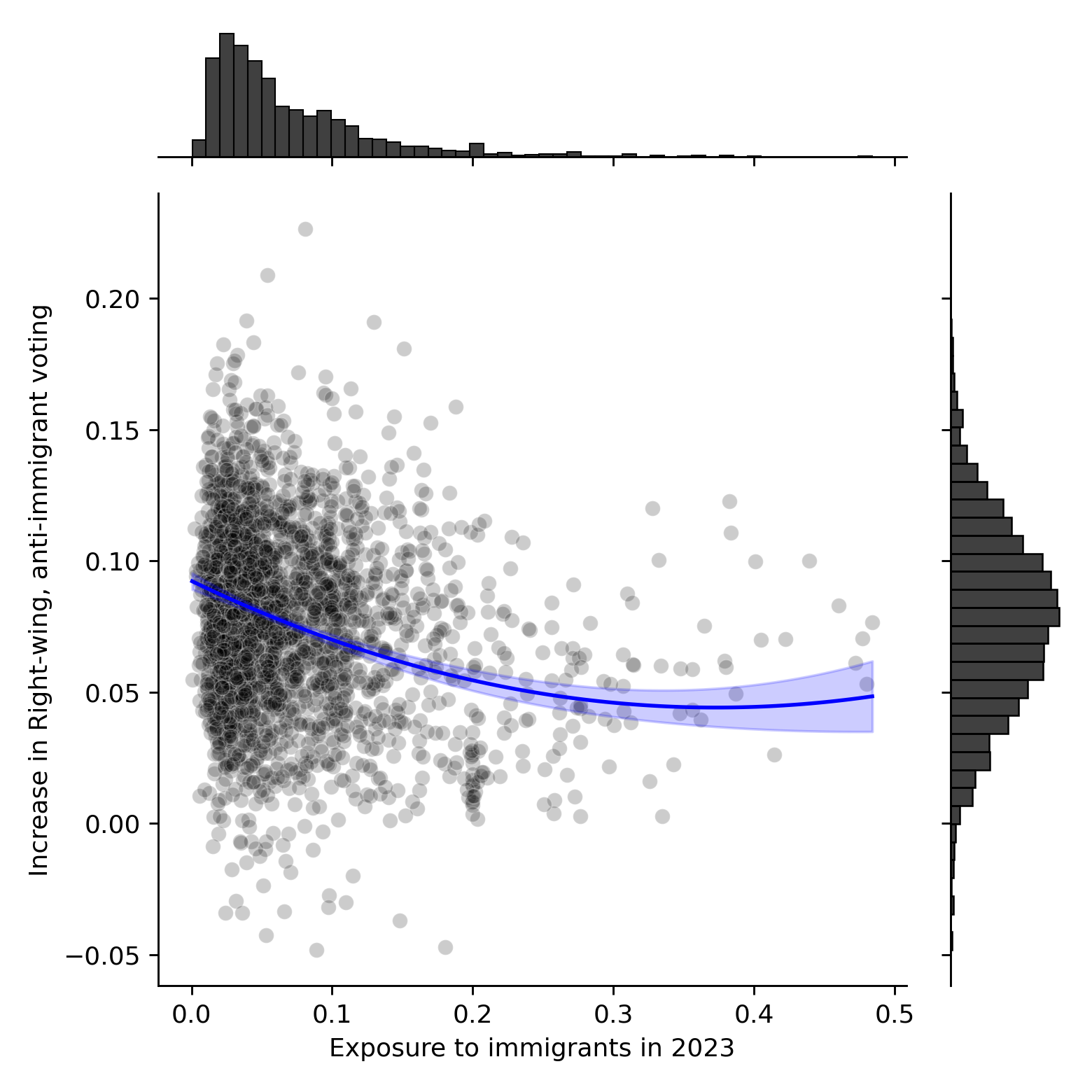}
   \caption{Quadratic (D2)}
   \label{fig:exp_diff_quad}
    \end{subfigure}%
    \caption{Exposure to immigrants in relationship to the increase in right-wing, anti-immigrant voting per municipality in models D1 and D2 in Table~\ref{tab:diff_vote} (in blue is the marginal effect as per the regression model, assuming the controls take their mean value).}
    \label{fig:Exp_delta_RW}
\end{figure}

In an ideal setting, we would investigate the relationship between changes in exposure and changes in right-wing vote share. Unfortunately, such a comparison is hindered by a change in the reporting of heritage in the Statistics Netherlands grid data.\footnote{This change was implemented in 2022, meaning that we cannot compare the exposure in 2021 to the exposure in 2023.} However, we can go as far as to use the election outcome of the Dutch general election of 2021. Here, we study the relationship between the exposure in 2023 and the change in right-wing vote share from 2021 to 2023. We do so using linear and quadratic models, specifically:
\begin{equation}
   y_{i,2023}- y_{i,2021} = \alpha E_{i,2023} + \bm{\beta}\cdot\texttt{Controls}_{i,2023} +\epsilon_i,
\end{equation}
and
\begin{equation}
  y_{i,2023}- y_{i,2021} = \alpha_1 E_{i,2023} + \alpha_2 E_{i,2023}^2 + \bm{\beta}\cdot\texttt{Controls}_{i,2023} +\epsilon_i,
\end{equation}
where $y_{i,a}$ is the right-wing vote share in neighbourhood $i$ in the year $a$. Similarly, all the right hand side variables are also indexed by year for completeness.
The results of the corresponding regression models are presented in Table~\ref{tab:diff_vote} for no fixed effects, provincial fixed effects, and COROP regions as fixed effects. For the models without regional fixed effects, the fit (assuming all variables except exposure take the mean value) is illustrated against scatter plots of the exposure in 2023 against the increase in right-wing vote share from 2021 to 2023 in Figure~\ref{fig:Exp_delta_RW}.

\begin{table}
\caption{Regression results of both the linear and quadratic model specifications (including insignificant coefficients) studying the change in right-wing vote share from 2021 to 2023 (positive implies an increase in right-wing vote share) against exposure and controls. The $k$ value used to compute exposure is $10\,000$. Standard errors in parentheses.}
\label{tab:diff_vote}
\begin{center}
\begin{tabular}{l|ll|ll|ll}
\hline
                            & \multicolumn{2}{c}{No Fixed effects} & \multicolumn{2}{c}{Province FE} & \multicolumn{2}{c}{COROP FE}  \\
                            & Linear & Quadratic & Linear & Quadratic & Linear & Quadratic  \\
                            & D1 & D2 & D3 & D4 & D5 & D6 \\

\hline
Exposure                  & -0.146***                     & -0.259***                 & -0.122***          & -0.206***                & -0.156***         & -0.243***              \\
                              & (0.014)                       & (0.028)                   & (0.013)            & (0.027)                  & (0.015)           & (0.027)                \\[4pt]
{Address density}   & 0.010***                      & 0.011***                  & 0.003**            & 0.004***                 & 0.003***          & 0.004***               \\
                              & (0.001)                       & (0.001)                   & (0.001)            & (0.001)                  & (0.001)           & (0.001)                \\[4pt]
{WOZ}   & -0.010***                     & -0.010***                 & -0.016***          & -0.016***                & -0.018***         & -0.018***              \\
                              & (0.001)                       & (0.001)                   & (0.001)            & (0.001)                  & (0.001)           & (0.001)                \\[4pt]
$\texttt{S}_{15,25}$             & -0.025                        & -0.035                    & 0.017              & 0.009                    & 0.019             & 0.010                  \\
                              & (0.023)                       & (0.023)                   & (0.022)            & (0.021)                  & (0.021)           & (0.021)                \\[4pt]
$\texttt{S}_{25,45}$             & 0.042***                      & 0.045***                  & 0.022*             & 0.024*                   & 0.008             & 0.012                  \\
                              & (0.015)                       & (0.015)                   & (0.013)            & (0.013)                  & (0.013)           & (0.013)                \\[4pt]
$\texttt{S}_{45,65}$             & 0.164***                      & 0.172***                  & 0.134***           & 0.140***                 & 0.125***          & 0.131***               \\
                              & (0.024)                       & (0.024)                   & (0.022)            & (0.022)                  & (0.021)           & (0.021)                \\[4pt]
Exposure$^2$ &                               & 0.346***                  &                    & 0.254***                 &                   & 0.273***               \\
                              &                               & (0.069)                   &                    & (0.064)                  &                   & (0.065)                \\[10pt]
R-squared                     & 0.134                         & 0.140                     & 0.349              & 0.352                    & 0.402             & 0.405                  \\
R-squared Adj.                & 0.132                         & 0.138                     & 0.345              & 0.348                    & 0.390             & 0.394                  \\
N                             & 2444                          & 2444                      & 2444               & 2444                     & 2444              & 2444                   \\
\hline
\multicolumn{3}{l}{*$p<.1$, **$p<.05$, ***$p<.01$}
\end{tabular}
\end{center}
\end{table}

From Table~\ref{tab:diff_vote} and Figure~\ref{fig:exp_diff_lin}, we see that, similar to our main findings, the change in right-wing vote share follows a contact theory type pattern. At low levels of exposure there is a broad band of possible increases in right-wing voting. As exposure increases we see less extreme increases in right-wing voting from 2021 to 2023. Note that there was a national increase; furthermore, this national increase was relatively consistent across the country. That is, most of the values for the difference between right-wing vote share from 2021 to 2023 are positive. The increase, however, was strongest in places with relatively low exposure to immigrants.

Following the findings of Kazmina \textit{et al.}~\cite{kazmina2024}, we also investigate a possible quadratic relationship. They found, using a LOWESS fit, that the relationship between exposure and anti-immigrant sentiment is not linear but exhibits a turning point. Before the turning point, more exposure is correlated to less anti-immigrant sentiment, while after the turning point, the relationship flips. We use a simple quadratic model specification (D2, D4, and D6 in Table~\ref{tab:diff_vote}, with D2 plotted in Figure~\ref{fig:exp_diff_quad}) studying exposure and the square thereof against the change in right-wing vote share. Here our results do align somewhat with those of Kazmina \textit{et al.}~\cite{kazmina2024}, though we are looking at the change in right-wing vote share rather than the vote share itself (used as a proxy for anti-immigrant sentiment). We see that the linear term has a significant negative relationship, as in the linear specification, but also that the quadratic term has a significant positive coefficient. This is true in all the quadratic models, including those with provincial or COROP regional fixed effects. Furthermore, the quadratic specification (regardless of including fixed effects) has a slightly better R-squared than the corresponding linear model, suggesting that the fit is improved by the inclusion of the quadratic term. Though the turning point occurs quite late in the data, it seems to indicate that there may be both contact and threat theory type behaviour, even on an aggregated scale. Exposure can get so high that the rate of increase in right-wing vote share picks up again.

Our findings provide a reconciliation of the results of Achard \textit{et al.}~\cite{Achard2024}, who find causal evidence for increased exposure leading to less anti-immigrant sentiment in the Netherlands, and Gravelle \textit{et al.}~\cite{Gravelle2021}, who illustrate that exposure to mosques \textit{with} minarets is related to greater right-wing voting in the Netherlands. The pattern in the 2023 election shows that, for the most part, higher exposure is related to less right-wing voting \textit{and} a lesser increase in right-wing voting (similar to what is found by, for instance, Achard \textit{et al.}~\cite{Achard2024}), but that this relationship flips around at a high enough exposure level, which may be the patterns that Gravelle \textit{et al.}~\cite{Gravelle2021} picked up in their study. 

\subsubsection{Biggest increase in right-wing vote share}

From Figure~\ref{fig:Exp_delta_RW}, we see that there are a few standout neighbourhoods. In particular, there are two that experienced an increase in right-wing voting of more than 20 percentage points. Furthermore, we see a couple of neighbourhoods with relatively high exposure (more than 0.3) alongside a relatively strong increase in right-wing vote share (more than 10 percentage points). These are likely helping the quadratic model achieve an improved adjusted R-squared compared to the linear model, and therefore warrant further attention.

The two neighbourhoods with an increase in right-wing vote share from 2021 to 2023 of more than 20 percentage points are:
\begin{itemize}
    \item Brouwhuis in Helmond, North Brabant: $RW_{23}=0.44$, $\Delta_{RW} = 0.226$, $E_{23} = 0.08$
    \item IJmuiden-Zuid in Velsen, North Holland: $RW_{23}=0.44$, $\Delta_{RW} = 0.209$, $E_{23} = 0.05$
\end{itemize}
Notably, both of these neighbourhoods had a very low measure of exposure. Perhaps more telling is that in Brouwhuis, in 2022, an old school building was converted to house up to 300 Ukrainian refugees~\cite{helmondVluchtelingenOekraine,ditishelmondBuurtavondOpvang}. Moreover, in IJmuiden, the cruise ship MS Silja Europa was used as an emergency housing centre for up to 1000 asylum seekers~\cite{VOBKade}. The Ukrainian refugees would not have been picked up by our measure, which defines immigrants as people born outside of the Netherlands with a non-European background, though 300 additional individuals are unlikely to shift the exposure measure. Rather than exposure to individuals, these cases illustrate that an acute concentration of immigrants with a clear visual reference (an old school building and a large cruise ship) may be the underlying mechanism that increases right-wing, anti-immigrant voting. It should also be noted that both of these neighbourhoods are in municipalities with a relatively high right-wing voter population across neighbourhoods.

As a related analysis, we conduct a robustness check of the models M2, M4, and M6, excluding the region containing Ter Apel in Westerwold, the neighbourhood with the central arrival point for asylum seekers in the Netherlands, in Appendix~\ref{app:TerApel}. In line with the above findings, in this robustness check we find that Ter Apel provides another example of regions where the dynamics are likely strongly influenced by a highly visible (and frequently reported on) asylum seekers centre. 

\subsubsection{Highly exposed neighbourhoods with high increase in right-wing vote share}
The higher exposure neighbourhoods ($E>0.3$), which also saw a significant increase, are four neighbourhoods in the south of The Hague: Moerwijk, Zuiderpark, Transvaalkwartier, and Bouwlust en Vrederust. The right-wing vote share in these neighbourhoods in 2023 was between 0.32, and 0.39 which is roughly 10 percentage points above the national average of 0.23. These neighbourhoods are characterised by relatively low wealth and high levels of social housing, with the greatest mean WOZ (housing value) among them being -1.19, more than a standard deviation away from the log-transformed global average (the log transformation and centering described in Appendix~\ref{app:vars}). Here, the likely (perceived) unfairness of the distribution of refugees may have played a bigger role than their actual presence. A populous faced with an influx of refugees who are now competing for the already scarce social housing may have triggered feelings of relative deprivation among the residents. 

To systematically analyse the increase in right-wing voting in `highly' exposed neighbourhoods, we restrict the data to neighbourhoods where $E\geq\theta$ for $\theta\in\{0.2, 0.25, 0.3\}$ and perform the same multiple regressions on this restricted dataset. Note that because there are relatively few neighbourhoods after the filtering, we no longer include provincial or COROP region fixed effects. The results are in Table~\ref{tab:diff_vote_filter}, which show that the only variable that is statistically significant at $p<0.01$ under all filters for `highly' exposed neighbourhoods is the WOZ value. From these results, we could say that if a neighbourhood is highly exposed, the response to this exposure is determined almost exclusively by our proxy for wealth. In poorer neighbourhoods, there is a bigger increase in right wing vote share, while in richer neighbourhoods, there is a smaller increase in right wing vote share from 2021 to 2023. We see that filtering on the `low' exposed neighbourhoods yields similar results to those for the full data (see D1 in Table~\ref{tab:diff_vote}).

\begin{table}
\caption{Regression results of the linear model (as in D1) on the data filtered to regions with `high', and `low' exposure (including insignificant coefficients) studying the change in right-wing vote share from 2021 to 2023 (positive implies an increase in right-wing vote share) against controls. The $k$ value used to compute exposure is $10\,000$. Standard errors in parentheses.}
\label{tab:diff_vote_filter}
\begin{center}
\begin{tabular}{l|ll|ll|ll}
\hline
 Data filtered to:     & $E_i\geqslant0.2$ &  $E_i <0.2$ & $E_i\geqslant0.25$  & $E_i<0.25$ & $E_i\geqslant0.3$  & $E_i<0.3$  \\
\hline
Exposure                    & 0.026     & -0.206*** & 0.047     & -0.205*** & 0.050     & -0.197***  \\
                            & (0.030)   & (0.021)   & (0.036)   & (0.019)   & (0.060)   & (0.016)    \\[4pt]
{Address density}    & -0.002    & 0.011***  & 0.003     & 0.011***  & 0.011     & 0.011***   \\
                            & (0.006)   & (0.001)   & (0.010)   & (0.001)   & (0.024)   & (0.001)    \\[4pt]
{WOZ}                & -0.017*** & -0.009*** & -0.018*** & -0.010*** & -0.021*** & -0.010***  \\
                            & (0.002)   & (0.001)   & (0.002)   & (0.001)   & (0.006)   & (0.001)    \\[4pt]
$\texttt{S}_{15,25}$        & -0.061    & -0.054**  & -0.125*   & -0.039    & -0.119    & -0.037     \\
                            & (0.056)   & (0.025)   & (0.069)   & (0.025)   & (0.167)   & (0.024)    \\[4pt]
$\texttt{S}_{25,45}$        & -0.047    & 0.073***  & -0.087*   & 0.056***  & -0.081    & 0.052***   \\
                            & (0.035)   & (0.017)   & (0.049)   & (0.016)   & (0.111)   & (0.015)    \\[4pt]
$\texttt{S}_{45,65}$        & 0.139**   & 0.168***  & 0.050     & 0.169***  & 0.064     & 0.166***   \\
                            & (0.070)   & (0.025)   & (0.081)   & (0.025)   & (0.130)   & (0.024)    \\[10pt]
R-squared                   & 0.655     & 0.113     & 0.637     & 0.127     & 0.540     & 0.139      \\
R-squared Adj.              & 0.638     & 0.110     & 0.603     & 0.124     & 0.445     & 0.137      \\
N                           & 127       & 2317      & 71        & 2373      & 36        & 2408       \\
\hline

\multicolumn{3}{l}{*$p<.1$, **$p<.05$, ***$p<.01$}
\end{tabular}
\end{center}
\end{table}

Fieldwork on public sentiment regarding newcomer reception centres by Hollander \textit{et al.}~\cite{Hollander2024} studied areas around seven reception centres for newcomers (asylum seekers and Ukrainian refugees) in the Hague. The share of respondents in the area around the Gulden Huis (Bouwlust en Vrederust), who report becoming more negative about ‘newcomers’ (24\%) exceeds the share who report becoming more positive (14\%). These figures are based on a relatively small number of respondents, so they should be interpreted as indicative rather than definitive. In five of the remaining six areas, there were greater positive shifts than negative shifts, while in one, these were equal. The results in Table~\ref{tab:diff_vote_filter} indicate that this may be due to feelings of relative deprivation in the population living near the Gulden Huis centre.

\section{Discussion}
The success of the right-wing anti-immigration Party for Freedom (PVV) in the 2023 national election has highlighted the importance of understanding what shapes voting behaviour. The PVV party programme for the 2025 national election opens with promises of working against asylum centres, against `mass-immigration', and against the Islamisation of the Netherlands~\cite{PVV2025Program}. Compared to the 2023 national election, they did not secure as much of the national vote share; though rounded to the nearest percentage point, they achieved a joint highest total vote share (17\%) as the winners D66. In light of the sustained inflow of asylum seekers and refugees from various armed conflicts around the world, alongside unforced migration, we see that it becomes increasingly important to understand how we can foster tolerance and understanding between natives and newcomers in the Netherlands. In this study, we have used granular grid data coupled with election data to uncover patterns in voting behaviour, building toward an understanding of when exposure can create space for intergroup connections and when these opportunities might get overshadowed by feelings of threat.

\subsection{Three behaviours at play}
Our investigation has highlighted three mechanisms for the interaction between exposure and right-wing voting behaviour. The first being the most suggested by the relationships in the data: Contact theory. Regions with greater exposure consistently show less right-wing anti-immigrant voting. This is suggested by the linear models for the relationship between exposure and right-wing voting in 2023, as well as the model for the increase in right-wing voting from 2021 to 2023 relative to exposure. For the latter, there is also a quadratic model that performs similarly well; though even there, for most levels of exposure, the pattern is that greater exposure is related to a lesser increase in right-wing voting from 2021 to 2023. When natives get the opportunity for meaningful contact, it can result in an increase in trust for the newcomers. {This aligns with the results of previous work on the Netherlands~\cite{vanHeerden2019,Achard2024,kazmina2024}.} However, these opportunities for interaction are not always taken nor do they always lead to warmer feelings toward the out-group.

The second mechanism our data suggest is that there is such a thing as \textit{too much} exposure, especially in times and places where there is economic hardship. We see that poorer regions (lower WOZ value) have greater right-wing voting. Beyond this, we see that the quadratic model between exposure and the increase in right-wing voting from 2021 to 2023 exhibits a turning point. After steadily having a lesser increase in right-wing voting in more exposed neighbourhoods, the relationship reverses, and greater exposure is related to greater increases. The presence of `too much' exposure switching the mechanism from contact to threat aligns with the work in the Netherlands~\cite{kazmina2024}, as well as work in France~\cite{Vertier2023}. From these highly exposed regions, we inspected the four with the greatest increase in right-wing vote share. All four of these were neighbourhoods of The Hague. Though perhaps more tellingly, all four had relatively low mean WOZ values, despite the Hague being a relatively rich city. Similarly, Arzheimer \textit{et al.}~\cite{Arzheimer2024} find that perceived decline increases the propensity for populist radical right support, and that the presence of immigrants is a strong predictor of perceived decline. %Aligning with the switch in our results is that they see that context like education levels moderate the relationship between the presence of immigrants and perceived decline.

The final mechanism we propose is related to \textit{visual reminders} coupled with acute increases in exposure. The two neighbourhoods with the overall greatest increase in right-wing voting from 2021 to 2023 were Brouwhuis in Helmond and IJmuiden-Zuid in Velsen. In both of these neighbourhoods, an impromptu solution was found to the necessity of housing asylum seekers and refugees over the years 2021-2023. In particular, IJmuiden featured a converted cruise ship that housed roughly 1,000 asylum seekers. The presence of such a large visible structure reminding residents of the presence of the newcomers may be shaping their beliefs more than their personal interactions (or lack thereof) with the group of newcomers. This aligns with Gravelle \textit{et al.}~\cite{Gravelle2021} in the Netherlands (and Valli \textit{et al.}~\cite{Valli2026} in Switzerland) who investigate the association between the presence of mosques and far-right voting, and find that it is especially the most visible elements (e.g., minarets) that drive the effects, rather than the simple presence of a mosque in the neighbourhood. Similarly, the idea that threat is not so much the presence of immigrants but rather something discernible and allusive aligns with results from Sweden, where support for the halo effect\footnote{—not the regions with high exposure, but rather regions close to high exposure to immigrants—was found to be associated with an increase in far-right voting.} rather than threat theory~\cite{rydgren2013contextual}. Finally, evidence suggests that while distant diversity is abstract, proximate but segregated micro-exposure—characterised by sudden visibility without meaningful interaction—can activate exclusionary shifts~\cite{enos2014causal}. This may be the driving force behind the change we observe at the lowest address densities, where the marginal effect of increased exposure is related to a greater right-wing vote share in our data.

The switch from a contact to a threat theory type relationship between dense urban regions and sparse rural regions also aligns with the findings of Garc\'ia-Mu\~noz~and~Milgram-Baleix~\cite{Munoz2021}, who show that exposure decreases the propensity for anti-immigrant sentiment among highly skilled people. The pattern for low-and medium-skilled individuals, however, is one of threat. The concentration of highly-skilled people in the urban regions is greater than in the sparse rural ones in the Netherlands.

The final two mechanisms are similar in that both feature opportunities for contact that are overshadowed. In the second, the personal situation of natives may get in the way of the contact, which is reported to decrease anti-immigrant sentiment in group contact theory. In the third, the opportunity for contact is overshadowed by the constant visual reminder of the presence of the newcomers. Constant reminders force attention on the topic, which, according to a set of opinion dynamics models, can lead to polarisation on that topic (see, for instance,~\cite{vanderMaas2020,Hoffstadt2025}). Such polarisation could explain the extreme rather than centrist voting behaviour.

So while our results align with the literature, we offer new insights into how these dynamics may have played out during the 2023 national election (other results are typically somewhat older), which was a special case related to the large win by the far-right PVV. Furthermore, the other studies in the Netherlands are typically on the individual scale, and while important, are limited by the number of responses. Our work uses election data that includes a much larger group (and avoids any issues related to people wanting to seem a certain way in a survey). We show that the same dynamics may be playing out at a large scale even when aggregating over individuals. Furthermore, this provides a promising direction for future work combining detailed census data and voting behaviour at an even greater granularity.

\subsection{Policy implications}
It may be tempting to convert existing structures into temporary housing for newcomers. This, however, can lead to sharp increases in exposure, which may in turn be related to a sharp increase in the anti-immigrant sentiment of the locals. Particular regions that may already be under stress due to housing shortages and low wealth may be particularly vulnerable to this phenomenon. Finally, when creating policy for housing newcomers, care should be taken to create space for and encourage meaningful interactions between natives and newcomers. One of the most obvious ways to do this is by communicating with the locals about the (future) presence of newcomer centres and ways in which they can be involved in the process of setting up and running the centres. 

Our results show patterns that are associated with group contact theory, which is typically studied as the relationship between exposure to immigrants and anti-immigrant sentiment (or, as we have done, right-wing voting). However, our results also indicate that patterns of threat theory may arise under certain conditions—possibly suggesting that the two theories can co-exist under specific circumstances to explain more complex situations. While we can measure the `actual' exposure, we cannot account for the \textit{perceived} exposure within individuals. The presence of visible reminders of the immigrant or refugee population, such as the cruise ship housing them (or mosques with minarets, see Valli \textit{et al.}~\cite{Valli2026}), may be more important than who is seen on the way to the grocer, or indeed may change the effect that the contact or exposure has on the native. Furthermore, the situation of the locals should be taken into account. This is illustrated by the increase in right-wing vote share in the highly exposed, relatively poor neighbourhoods of The Hague in our data, as well as the work by Arzheimer \textit{et al.}~\cite{Arzheimer2024} which highlights the links between perceived decline, the presence of immigrants, and populist right-wing voting. If the local population feels neglected and forgotten, they may not respond warmly to an influx of immigrants or refugees, as this could trigger feelings of relative deprivation (see for instance~\cite{Gheorghiu2022}).

The specifics of how immigration is handled may be incredibly important to the analysis. For instance, while causal evidence for contact theory was found in the case of refugees in the Netherlands from 2011-2016~\cite{Achard2024}, Hangartner \textit{et al.}~\cite{hangartner2019exposure} find causal evidence for group threat theory in the Aegean Sea area. Similarly, a longitudinal study of Japan and the UK shows more likeness with threat theory~\cite{Igarashi2021}. The way in which national governments handle refugee situations and immigration in general may play a large role in determining how the exposure affects the native population.

\subsection{Limitations}
Life-time exposure is not taken into account in any of our models. Individuals who spend 10 years living in highly exposed Amsterdam and then return to their rural roots likely have different voting behaviour than their neighbours who stayed their whole lives in the village.

The setup using right-wing votes as a proxy allows us to investigate the entire voting population of the Netherlands; however, it forces us to use aggregated measures. Care has been taken to compute the aggregated measure as the average of that measure experienced by the population in the unit studied. Inherently, by studying population aggregates, we run the risk of ecological fallacy. This also means that individual patterns may be missed, where a part of the population responds in one way to exposure, while another part responds in the opposite way. For our purposes, we are grateful for the previous work, in particular that of Achard~\textit{et al.}~\cite{Achard2024}, who performed a careful study on the individual level in the Netherlands based on roughly 5,000 respondents in the LISS panel during the period 2011-2016. We thus see our own work as a natural extension of this switching out causal identification capability for range, studying the entire country within the context of the more recent 2023 election.

The data we study are already aggregated at a scale of 500 by 500 metres and subsequently at the level of neighbourhood and municipality. This means we could miss dynamics arising at a scale smaller than this. Pockets of increased right-wing voting within a context of the opposite may go unnoticed. Furthermore, though the proxy we use for anti-immigrant sentiment is supported by the LISS data and commonly used in the literature, there is a chance that voting behaviour is dictated by factors other than anti-immigrant sentiment.

\addcontentsline{toc}{section}{References}

\bibliographystyle{ieeetr}
\bibliography{ref.bib}

%%%%%%

\appendix
\section{Non-colinearity of control variables}\label{app:vars}

\begin{figure}[htb]
    \centering
       \begin{subfigure}{0.5\textwidth}
        \includegraphics[width=0.95\textwidth]{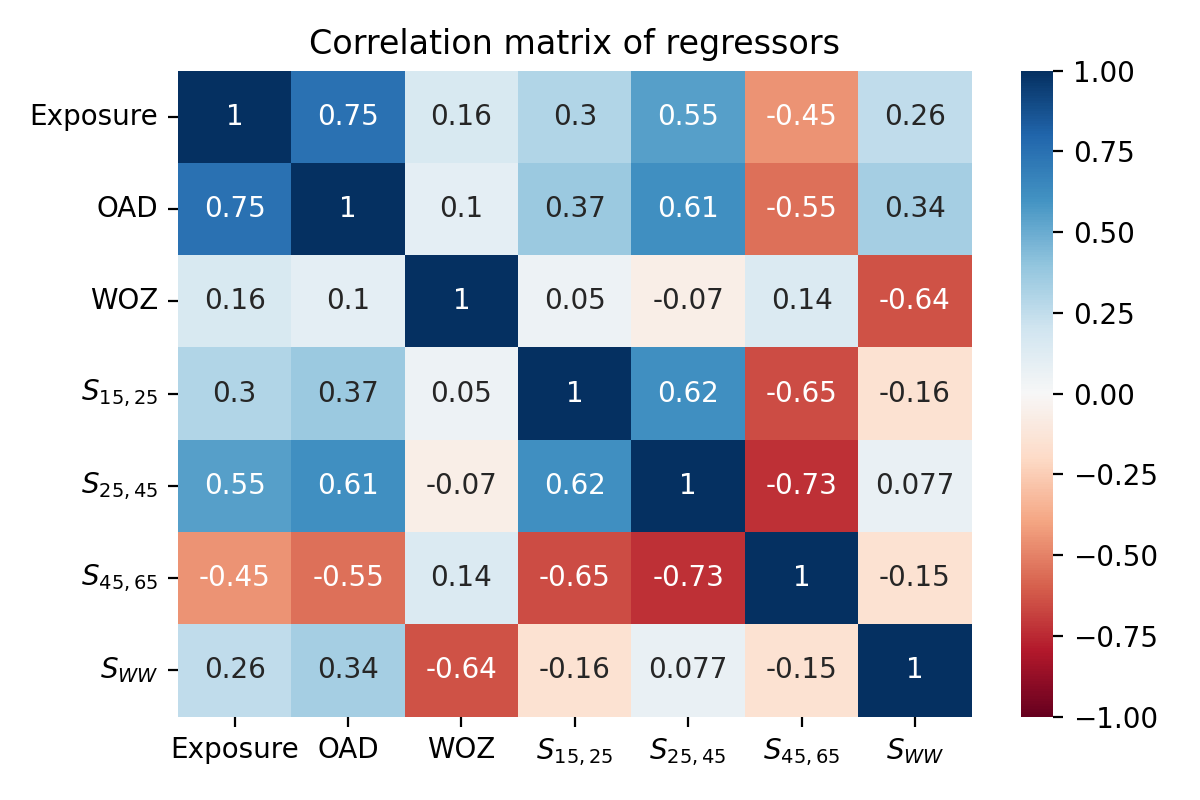}
        \caption{Municipality}
    \end{subfigure}%
    \begin{subfigure}{0.5\textwidth}
        \includegraphics[width=0.95\textwidth]{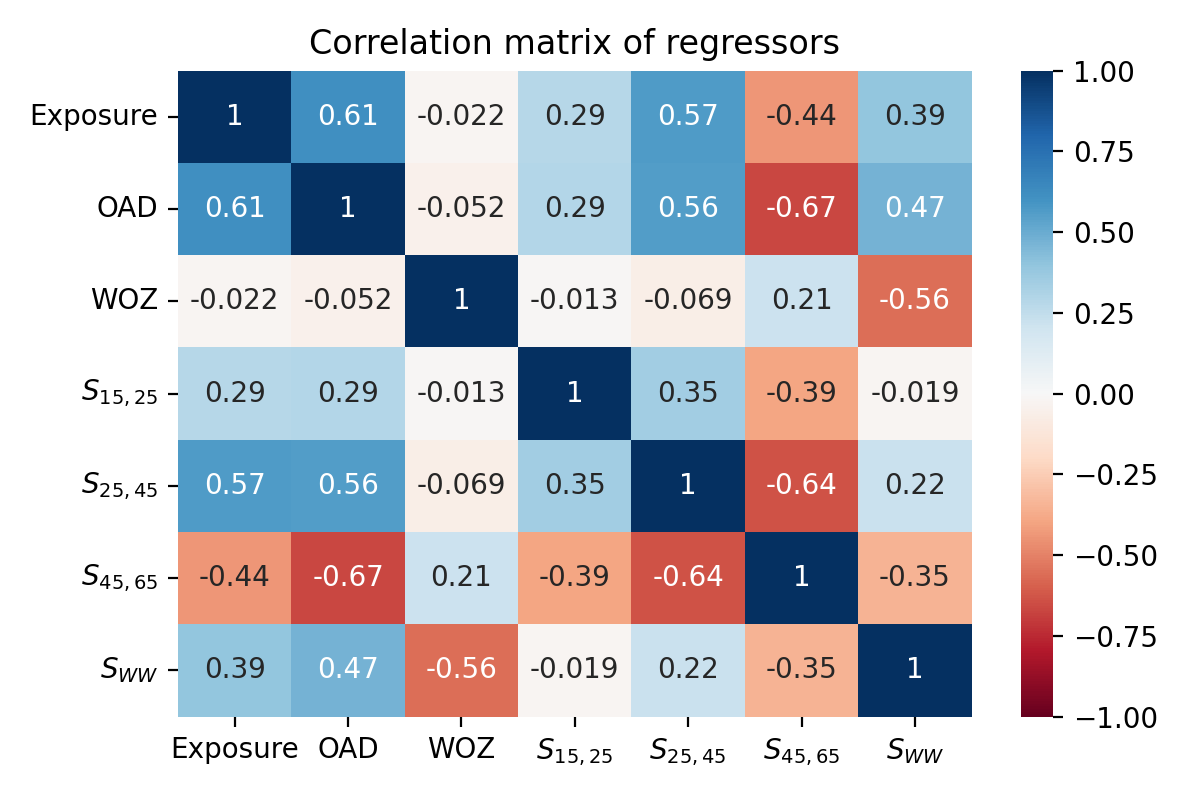}
        \caption{Neighbourhood}
    \end{subfigure}
    \caption{Correlation matrices of the regressors on the scale of municipality and neighbourhood. OAD: Address density, WOZ: Home value, $S_{WW}$: Share receiving benefits.}
    \label{fig:corr_mat}
\end{figure}

The correlation among the regressors (displayed in the correlation matrix in Figure~\ref{fig:corr_mat}), is not high enough to cause concern related to colinearity. The greatest (anti-)correlation is -0.73 between $\texttt{S}_{25,45}$ and $\texttt{S}_{45,65}$ on the scale of municipalities. For neighbourhoods the greatest (anti-)correlation is -0.67 between $\texttt{S}_{45,65}$ and address density (OAD). These values are reasonable and sensible, folks in the age range 45--65 do not appear to be living in the high density urban areas. Similarly, it is also reasonable that the home value (WOZ) is somewhat anti-correlated with the share of individuals receiving government benefits ($S_{WW}$). 

In order to check for multicollinearity we inspect the variance inflation factors (VIF) for the main model specifications M2 and M6 from Table~\ref{tab:baseModelRegr}. These are tabulated in Table~\ref{tab:VIF_basic}. None of these values are so high as to require correction.

\begin{table}[h]
    \caption{Variance inflation factors at the Neighbourhood scale.}
    \label{tab:VIF_basic}
    \centering
    \begin{tabular}{l|ll}
            Variable  &   \multicolumn{1}{c}{VIF M2} & \multicolumn{1}{c}{VIF M6} \\
              &        &  \multicolumn{1}{c}{(COROP FE)} \\
            \hline
           Exposure  &  1.852 & 2.688\\[4pt]
            {Address density}  &  2.384 & 3.009\\[4pt]
            {WOZ value} &   1.073 & 2.218\\[4pt]
            $\texttt{S}_{15,25}$  &  1.224 & 1.356\\[4pt]
            $\texttt{S}_{25,45}$    &  2.054 & 2.198\\[4pt]
            $\texttt{S}_{45,65}$    & 2.517 & 2.622\\
            \hline
    \end{tabular}

\end{table}

\begin{figure}[htb]
    \centering
       \begin{subfigure}{0.9\textwidth}
        \includegraphics[width=0.95\textwidth]{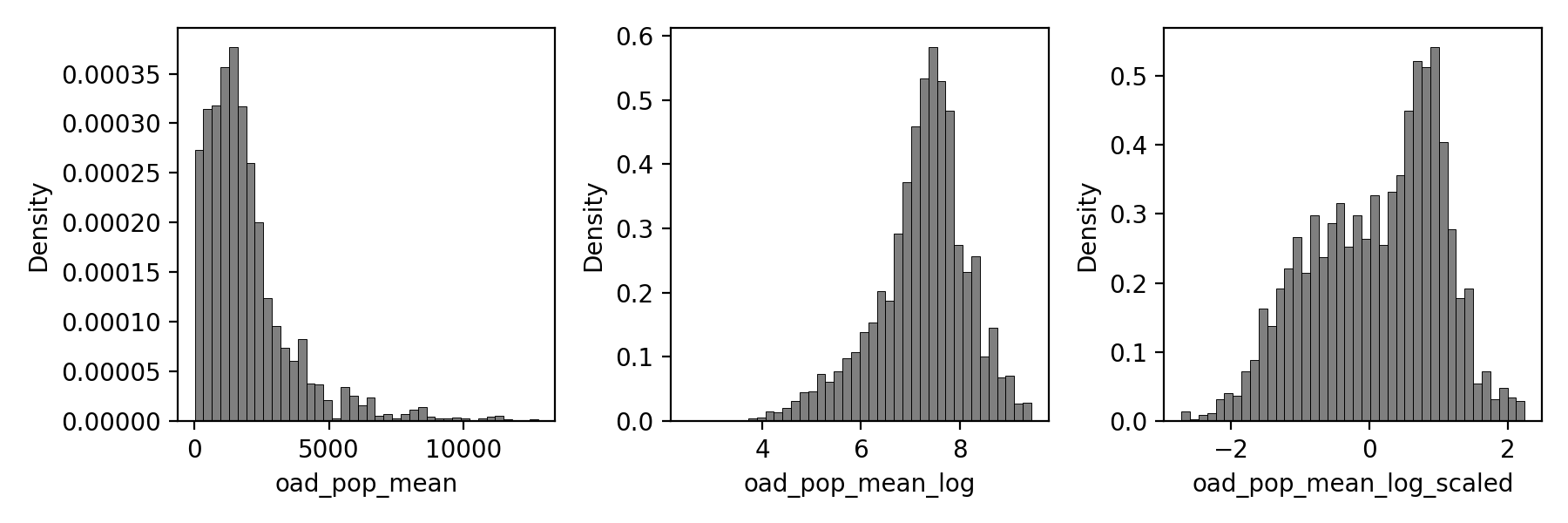}
        \caption{\texttt{Address density}}
    \end{subfigure}\\
    \begin{subfigure}{0.9\textwidth}
        \includegraphics[width=0.95\textwidth]{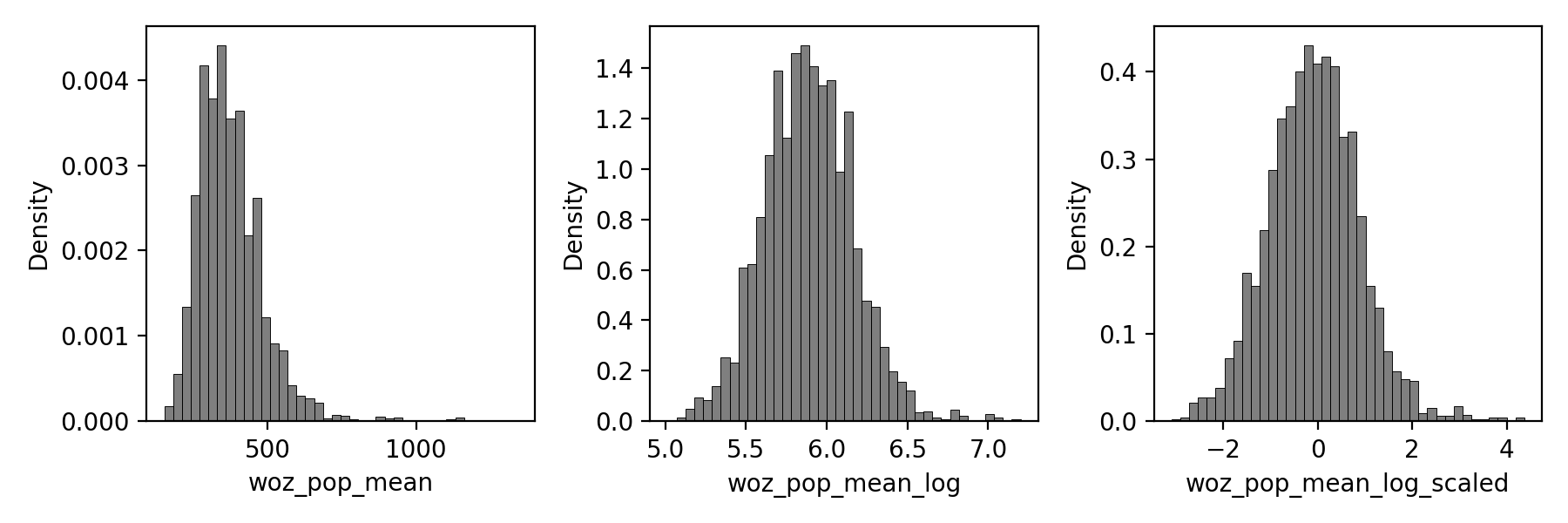}
        \caption{\texttt{WOZ value}}
    \end{subfigure}
    \caption{Transformation applied to the address density and the WOZ value.}
    \label{fig:log_scaling}
\end{figure}

For address density and the WOZ home value variables, we apply a log transformation as well as a centering. The effect this has on the variable is depicted in Figure~\ref{fig:log_scaling} plotting the histograms for these variables on the neighbourhood scale for the raw data, the log-transformed data, and finally the centred data. The naming convention indicates that these are the weighted mean values of the neighbourhood where the weighting is by population.

\section{Robustness checks}

\subsection{Different value of $k$}\label{app:diff_k}
To test the sensitivity on the value of $k$ used to compute the exposure, we run regressions on the neighbourhood scale for the exposure calculated using $k\in\{100, 1\,000, 5\,000\}$. These results are shown in Table~\ref{tab:robust_k_reg} and are remarkably similar to those of Table~\ref{tab:baseModelRegr} (the column on the neighbourhood scale). All coefficients have the same sign, and the coefficients for the exposure term are the same when rounded to the first digit after the decimal.

\begin{table}[htb!]
\caption{Regression results of the models under exposure computed with $k\in\{100, 1\,000, 5\,000\}$ (including insignificant coefficients). All of these models are run at the neighbourhood scale. Standard errors in parentheses.}
\label{tab:robust_k_reg}
\begin{center}
\begin{tabular}{l|lll}
\hline
                            & No FE & Province FE  & COROP FE \\
                            & M2 & M4 & M6 \\
\hline
% Exposure ($k=100$)         & -0.276***           & -0.319*** & -0.370***  \\

% \texttt{Address density} & 0.019***            & 0.002     & 0.001      \\

% \texttt{WOZ value} & -0.051***           & -0.064*** & -0.067***  \\
 
% $\texttt{S}_{15,25}$           & -0.273***           & -0.180*** & -0.151***  \\

% $\texttt{S}_{25,45}$          & 0.089***            & 0.045*    & 0.033      \\

% $\texttt{S}_{45,65}$           & 0.539***            & 0.489***  & 0.482***   \\

% R-squared                   & 0.368               & 0.529     & 0.592      \\
% R-squared Adj.              & 0.366               & 0.526     & 0.586      \\
%  N                           & 2816                & 2816      & 2816       \\
Exposure ($k=100$)        & -0.276***           & -0.319*** & -0.370***  \\
                            & (0.024)             & (0.023)   & (0.027)    \\[10pt]
R-squared                   & 0.368               & 0.529     & 0.592      \\
R-squared Adj.              & 0.366               & 0.526     & 0.586      \\[4pt]
\hline
Exposure ($k=1\,000$)       & -0.278***           & -0.320*** & -0.370***  \\
                            & (0.024)             & (0.023)   & (0.028)    \\[10pt]
R-squared                   & 0.368               & 0.529     & 0.592      \\
R-squared Adj.              & 0.367               & 0.526     & 0.586      \\[4pt]
\hline
Exposure ($k=5\,000$)          & -0.295***           & -0.342*** & -0.405***  \\
                            & (0.024)             & (0.023)   & (0.029)    \\[10pt]
R-squared                   & 0.370               & 0.531     & 0.595      \\
R-squared Adj.              & 0.368               & 0.528     & 0.589      \\[10pt]
N                           & 2816                & 2816      & 2816       \\
\hline
\multicolumn{3}{l}{*$p<.1$, **$p<.05$, ***$p<.01$}
\end{tabular}
\end{center}

\end{table}
The similarity between these two regression results, in addition to the shrinkage illustrated in Figure~\ref{fig:exposure_k_robust} indicates that the choice of $k$ does not strongly influence the result of the analysis. It should be acknowledged that this may also be related to the aggregated, grid nature of our analysis. If we construct the same measures using the precise locations of all residents in the Netherlands, the convergence of the exposure measure may possibly occur at higher values of $k$.

\subsection{Historic municipality voting as neighbourhood control}\label{app:historyMuni}

The 2021 CBS grid data from which we draw most of our control and explanatory variables are not comparable to the 2023 data. In particular, the reporting groups for heritage are redefined without clear mapping from one to the other. Previously, heritage was reported as western or non-western for individuals born outside the Netherlands; this has shifted to country of birth being inside or outside of the European Union. The voting data, however, is reported per municipality and per voting location for the national election that took place in 2021.

We test the robustness of our neighbourhood level results by including the municipality voting outcome to which the neighbourhood belongs as a control. In this way, we hope to account for the ideology and voting behaviour present in a small region (municipality) and to see if this explains the 2023 neighbourhood level election result. 

\begin{table}
\caption{Regression results for the robustness check using the 2021 national election results (on Municipal level) as control for the neighbourhood level analysis. The models listed are without, with province, and with COROP region fixed effects (FE). Standard errors in parentheses.}
\label{tab:muni21}
\begin{center}
\begin{tabular}{l|lll}
\hline
                            & No FE 		& Province FE      & COROP FE        \\
                            & H1		& H2		   & H3                 \\
\hline
Exposure		    & -0.141***         & -0.180***        & -0.255***        \\
                            & (0.021)           & (0.022)          & (0.026)          \\[4pt]
{Address density}    & 0.016***          & 0.005**          & 0.005**          \\
                            & (0.002)           & (0.002)          & (0.002)          \\[4pt]
{WOZ} value 	    & -0.028***         & -0.042***        & -0.050***        \\
                            & (0.001)           & (0.002)          & (0.002)          \\[4pt]
$\texttt{S}_{15,25}$        & -0.093***         & -0.082**         & -0.081**         \\
                            & (0.035)           & (0.035)          & (0.036)          \\[4pt]
$\texttt{S}_{25,45}$        & 0.133***          & 0.096***         & 0.071***         \\
                            & (0.022)           & (0.021)          & (0.022)          \\[4pt]
$\texttt{S}_{45,65}$        & 0.399***          & 0.393***         & 0.407***         \\
                            & (0.044)           & (0.045)          & (0.046)          \\[4pt]
$RW_{21}$                   & 1.017***          & 0.850***         & 0.791***         \\
                            & (0.035)           & (0.040)          & (0.049)          \\[10pt]

R-squared                   & 0.600             & 0.655            & 0.675            \\
R-squared Adj.              & 0.599             & 0.653            & 0.669            \\
N                           & 2816              & 2816             & 2816             \\
\hline
\multicolumn{3}{l}{*$p<.1$, **$p<.05$, ***$p<.01$}
\end{tabular}
\end{center}
%*$p<.1$, **$p<.05$, ***$p<.01$
\end{table}

From the results of this robustness check in Table~\ref{tab:muni21} we see that the level of right-wing voting in the municipality of each neighbourhood in 2021 has a very strong relationship with the right-wing voting of the neighbourhood in 2023. This is to be expected, though we see the strength of this relationship decrease as we include provincial fixed effects to the model, and again when we use COROP regions to capture fixed effects. 

Most important for our analysis is that the relationship between exposure and right-wing voting remains negative and significant. Thus, our results from the main text are robust against the inclusion of historical regional voting as a control. The strength of the relationship decreased somewhat compared to the original models (M2, M4, and M6 in Table~\ref{tab:baseModelRegr}), roughly halving in magnitude.

\subsection{Robustness against definition of right-wing}\label{app:PVV}
Considering that $PVV$ is consistently getting a greater number of votes than the other parties in $\mathcal{RW}$, we perform a robustness check using only the portion of votes for $PVV$ as the proxy for anti-immigration sentiment. The results of these regressions are in Table~\ref{tab:PVV_rob}. The results are robust to this specification.

\begin{table}[htb!]
\caption{Regression results of the models regressing against only PVV votes (including insignificant coefficients). All of these models are run at the neighbourhood scale.}
\label{tab:PVV_rob}
\begin{center}
\begin{tabular}{l|lll}
\hline
                            & No FE & Province FE  & COROP FE \\
                             & M23 & M24 & M25 \\
\hline
Exposure                    & -0.318***                   & -0.343***        & -0.423***        \\
                            & (0.023)                     & (0.023)          & (0.028)          \\[4pt]
{Address density}    & 0.020***                    & 0.004*           & 0.003            \\
                            & (0.002)                     & (0.002)          & (0.002)          \\[4pt]
{WOZ value}          & -0.046***                   & -0.058***        & -0.062***        \\
                            & (0.001)                     & (0.001)          & (0.002)          \\[4pt]
 
$\texttt{S}_{15,25}$        & -0.248***                   & -0.161***        & -0.126***        \\
                            & (0.040)                     & (0.037)          & (0.036)          \\[4pt]

$\texttt{S}_{25,45}$        & 0.076***                    & 0.035            & 0.020            \\
                            & (0.026)                     & (0.023)          & (0.024)          \\[4pt]

$\texttt{S}_{45,65}$        & 0.512***                    & 0.464***         & 0.459***         \\
                            & (0.054)                     & (0.052)          & (0.052)          \\[10pt]

R-squared                   & 0.364                  & 0.530       & 0.596       \\
R-squared Adj.              & 0.362                  & 0.528       & 0.589       \\
 N                           & 2816                & 2816      & 2816       \\
\hline
\multicolumn{3}{l}{*$p<.1$, **$p<.05$, ***$p<.01$}
% \hline
% Intercept                   & 0.100***               & 0.049**     & 0.107***    \\
%                             & (0.023)                & (0.023)     & (0.025)     \\
% pers\_weigh\_exp10000       & -0.318***              & -0.343***   & -0.423***   \\
%                             & (0.023)                & (0.023)     & (0.028)     \\
% oad\_pop\_mean\_log\_scaled & 0.020***               & 0.004*      & 0.003       \\
%                             & (0.002)                & (0.002)     & (0.002)     \\
% woz\_pop\_mean\_log\_scaled & -0.046***              & -0.058***   & -0.062***   \\
%                             & (0.001)                & (0.001)     & (0.002)     \\
% share\_15\_to\_25           & -0.248***              & -0.161***   & -0.126***   \\
%                             & (0.040)                & (0.037)     & (0.036)     \\
% share\_25\_to\_45           & 0.076***               & 0.035       & 0.020       \\
%                             & (0.026)                & (0.023)     & (0.024)     \\
% share\_45\_to\_65           & 0.512***               & 0.464***    & 0.459***    \\
% R-squared                   & 0.364                  & 0.530       & 0.596       \\
% R-squared Adj.              & 0.362                  & 0.528       & 0.589       \\
% N                           & 2816                   & 2816        & 2816        \\
% \hline
\end{tabular}
\end{center}
%*$p<.1$, **$p<.05$, ***$p<.01$
\end{table}

\subsection{Removing Ter Apel}\label{app:TerApel}
Ter Apel, a neighbourhood in Westerwolde, Groningen, houses an asylum seekers centre at which almost all seeking refuge in the Netherlands have to register as such upon arrival. To test the robustness of our approach, we run the main model specification excluding first the whole province of Groningen (112 neighbourhoods) and then the municipality of Westerwolde (8 neighbourhoods). The results of this robustness check are in Table~\ref{tab:robust_ter_apel}. 

\begin{table}[htb!]
\caption{Regression results on the neighbourhood scale when removing regions (province Groningen, and municipality Westerwolde) including the village Ter Apel from the analysis. The $k$ value used to compute exposure is $10\,000$. Standard errors in parentheses.}
\label{tab:robust_ter_apel}
\begin{center}
\resizebox{\textwidth}{!}{
\begin{tabular}{l|ll|ll|ll}
\hline

                            & \multicolumn{2}{c}{No Fixed effects} & \multicolumn{2}{c}{Province FE} & \multicolumn{2}{c}{COROP FE}  \\
                            & -Groningen & -Westerwolde & -Groningen & -Westerwolde & -Groningen & -Westerwolde  \\
                            & TA1 & TA2 & TA3 & TA4 & TA5 & TA6 \\

\hline
Exposure       & -0.304***    & -0.304***    & -0.353***    & -0.356***    & -0.431***     & -0.427***      \\
                            & (0.024)      & (0.024)      & (0.024)      & (0.024)      & (0.030)       & (0.029)        \\[4pt]
{Address density}    & 0.019***     & 0.021***     & 0.003        & 0.004*       & 0.002         & 0.004          \\
                            & (0.003)      & (0.003)      & (0.002)      & (0.002)      & (0.003)       & (0.003)        \\[4pt]
{WOZ value}          & -0.053***    & -0.050***    & -0.064***    & -0.063***    & -0.068***     & -0.067***      \\
                            & (0.002)      & (0.001)      & (0.002)      & (0.002)      & (0.002)       & (0.002)        \\[4pt]
$\texttt{S}_{15,25}$        & -0.253***    & -0.264***    & -0.169***    & -0.164***    & -0.146***     & -0.134***      \\
                            & (0.044)      & (0.042)      & (0.042)      & (0.039)      & (0.041)       & (0.038)        \\[4pt]
$\texttt{S}_{25,45}$        & 0.096***     & 0.096***     & 0.055**      & 0.054**      & 0.036         & 0.037          \\
                            & (0.028)      & (0.028)      & (0.025)      & (0.025)      & (0.026)       & (0.025)        \\[4pt]
$\texttt{S}_{45,65}$        & 0.549***     & 0.549***     & 0.499***     & 0.502***     & 0.497***      & 0.496***       \\
                            & (0.059)      & (0.058)      & (0.056)      & (0.055)      & (0.057)       & (0.056)        \\[10pt]
% Fixed Effects               & None         & None         & Prov.        & Prov.        & COROP         & COROP          \\
% Removed                     & Groningen    & Westerwolde  & Gronignen    & Westerwolde  & Groningen     & Westerwolde   \\
R-squared                   & 0.387        & 0.369        & 0.540        & 0.534        & 0.602         & 0.597          \\
R-squared Adj.              & 0.386        & 0.368        & 0.537        & 0.531        & 0.595         & 0.591          \\
N                           & 2704         & 2808         & 2704         & 2808         & 2704          & 2808           \\
\hline
\multicolumn{3}{l}{*$p<.1$, **$p<.05$, ***$p<.01$}
\end{tabular}}
\end{center}
\end{table}

\begin{figure}[htb!]
    \centering
       \begin{subfigure}{0.4\textwidth}
    \includegraphics[width=0.8\linewidth]{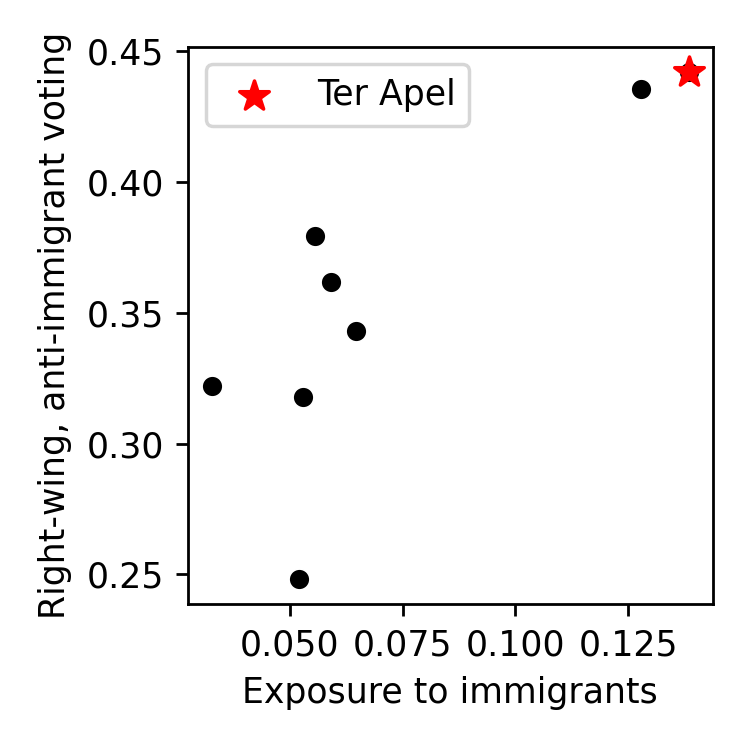}
    \caption{Scatter plot}
    \label{fig:TerApelScatter}
    \end{subfigure}%
    \begin{subfigure}{0.4\textwidth}
    \includegraphics[width=0.8\linewidth]{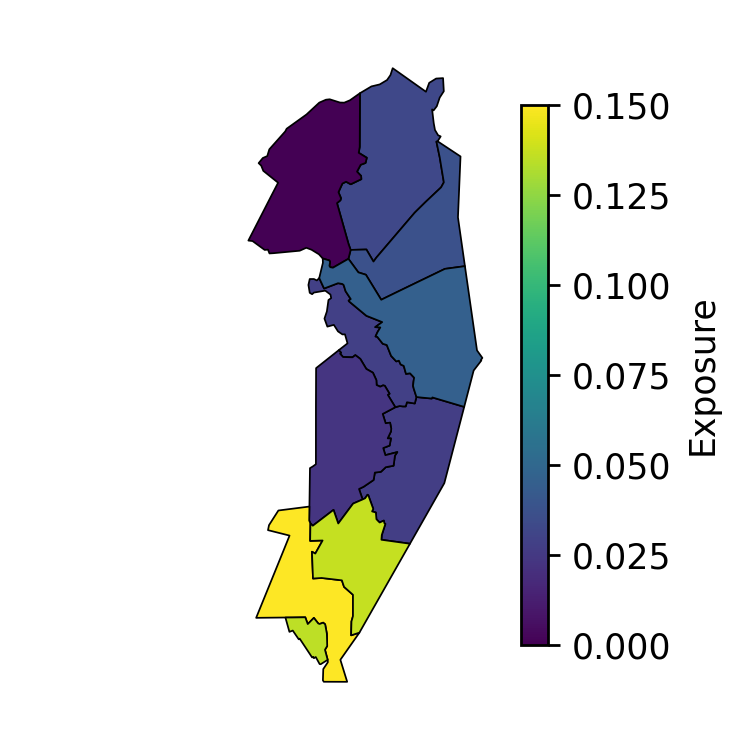}
    \caption{Map of Westerwolde}
    \label{fig:TerApelMap}
    \end{subfigure}%
    \caption{Exposure and right-wing voting in Westerwolde, the municipality housing Ter Apel. (a) is a scatter plot of exposure and right-wing voting, and (b) is a map illustrates the layout of this municipality and exposure of its neighbourhoods.}
    \label{fig:TerApelRobust}
\end{figure}

We see that the main results and patterns of Table~\ref{tab:baseModelRegr} are robust to the exclusion of Ter Apel. On closer inspection, we see that the coefficient of exposure increased in magnitude (became more negative) compared to the models including Ter Apel at the neighbourhood level (M2, M4, and M6). In addition, removing Ter Apel has allowed for a slightly improved fit (better adjusted R-squared), which implies that the situation in Ter Apel contradicts the main pattern of more exposure being related to less right-wing voting. This is confirmed by the plot of exposure and right-wing voting in the municipality of Westerwolde in Figure~\ref{fig:TerApelRobust}.

\section{Additional analyses}

\subsection{Quadratic exposure terms}\label{app:quad_exp}
\label{app:quad}
In Kazmina \textit{et al.}~\cite{kazmina2024} and Achard \textit{et al.}~\cite{Achard2024} it is reported that exposure has the strongest contact effect when it is not `too' great. Here, we test whether such behaviour is captured in our aggregated data by running regressions including a quadratic term for exposure. The resulting coefficients are presented in Table~\ref{tab:quad_exp}. The quadratic exposure term has a significant coefficient in all specifications. However, it is negative, meaning that the relationship does not weaken or turn around for higher exposure, but rather gets stronger. This is in contrast to the u-shape reported by Kazmina \textit{et al.}~\cite{kazmina2024}.

\begin{table}[htb]
\caption{Ordinary least squares regression on models including a quadratic term for exposure to immigrants. Three models are included, one without any regional fixed effects, one with provincial fixed effects and finally one with COROP fixed effects.}
\label{tab:quad_exp}
\begin{center}
\begin{tabular}{l|lll}
\hline
                            & No FE & Prov.\,FE & COROP FE  \\
                             & M20 & M21 & M22\\
\hline
Exposure                      & -0.148***    & -0.165***        & -0.220***          \\
                              & (0.057)      & (0.053)          & (0.052)            \\[4pt]
Exposure$^2$                  & -0.480***    & -0.575***        & -0.654***          \\
                              & (0.136)      & (0.129)          & (0.128)            \\[4pt]
{Address density}      & 0.019***     & 0.002            & 0.001              \\
                              & (0.003)      & (0.003)          & (0.003)            \\[4pt]
{WOZ value}            & -0.050***    & -0.063***        & -0.067***          \\
                              & (0.001)      & (0.002)          & (0.002)            \\[4pt]
$\texttt{S}_{15,25}$          & -0.251***    & -0.151***        & -0.113***          \\
                              & (0.042)      & (0.040)          & (0.039)            \\[4pt]
$\texttt{S}_{25,45}$          & 0.089***     & 0.045*           & 0.029              \\
                              & (0.028)      & (0.025)          & (0.025)            \\[4pt]
$\texttt{S}_{45,65}$          & 0.542***     & 0.492***         & 0.487***           \\
                              & (0.058)      & (0.055)          & (0.055)            \\[10pt]
R-squared                     & 0.371        & 0.533            & 0.599              \\
R-squared Adj.                & 0.370        & 0.530            & 0.592              \\
N                             & 2816         & 2816             & 2816               \\
\hline
\multicolumn{3}{l}{*$p<.1$, **$p<.05$, ***$p<.01$}
\end{tabular}
\end{center}

\end{table}

\subsection{Interaction with a population size categorical variable}
\begin{figure}[hb!]
    \centering
       \begin{subfigure}{0.5\textwidth}
    \includegraphics[width=0.9\linewidth]{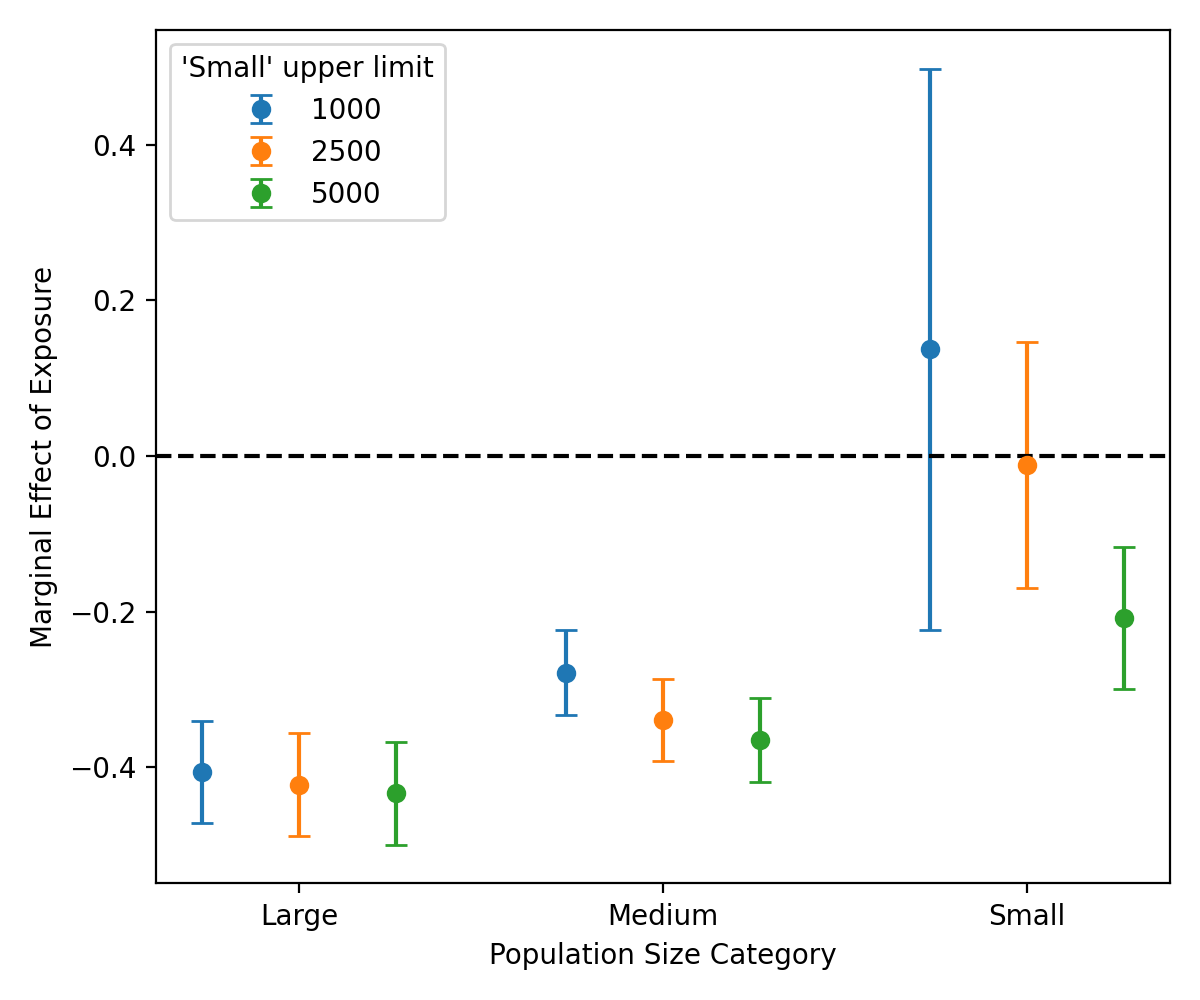}
    \caption{Provincial fixed effects}
    \label{fig:coeff_pop_catProv}
    \end{subfigure}%
    \begin{subfigure}{0.5\textwidth}
    \includegraphics[width=0.9\linewidth]{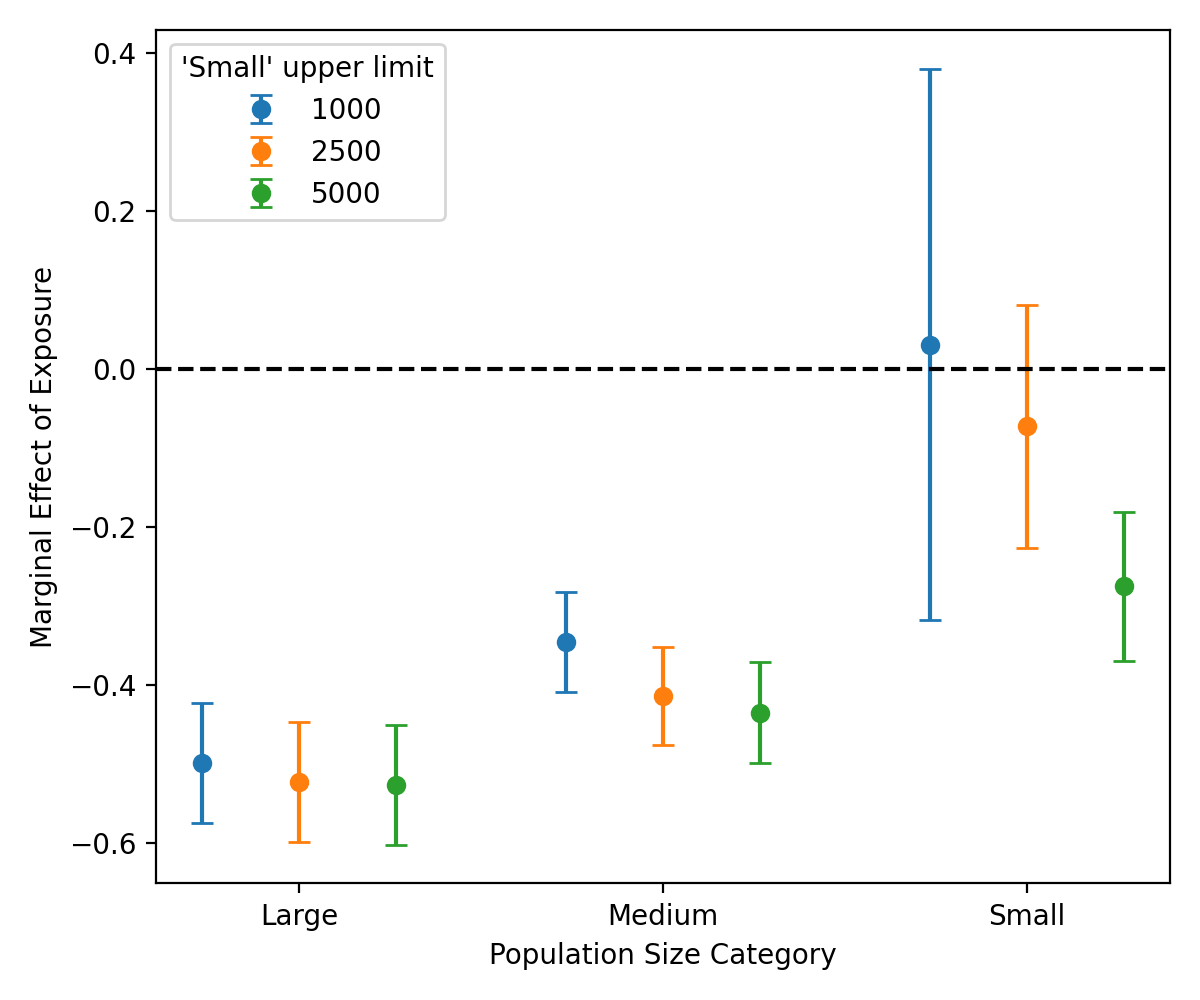}
    \caption{COROP fixed effects}
    \label{fig:coeff_pop_catCOROP}
    \end{subfigure}%
    \caption{The marginal effect of exposure under interaction with population size categories. In (a) the province accounts for fixed effects while in (b) this is handled by COROP regions.}
    \label{fig:coeff_pop_cat}
\end{figure}

In a meta-analysis, Kaufmann and Goodwin~\cite{Kaufmann2018} find that the population size of the unit of measurement matters for the result of whether the exposure results in contact or threat. They observe that for small units (smaller than 1,000 to 5,000), the threat is most likely to be observed, followed by contact for medium sized units (up to 10,000). Finally, for large population sizes, they state that threat is once again the dominant finding in the literature. To test this in our own data, we create a new categorical variable that classifies each neighbourhood as `small', `medium', or `large' depending on the total population in the neighbourhood. We vary the upper limit of the small category for robustness. The resulting marginal effect of exposure under an interaction with this categorical variable is plotted in Figure~\ref{fig:coeff_pop_cat}. In contrast to their results, we see that the association with exposure becomes more strongly negative for increasing group sizes. Threat theory is not convincingly supported for any of the groups. Furthermore, the effect becomes stronger for the largest group size rather than weaker, as predicted by their meta-analysis. The adjusted $R^2$ for these models is around 0.53 for all specifications in the small category. Thus, they do not outperform the model with interactions with the address density and home value and only slightly outperform the main specification M6 in Table~\ref{tab:baseModelRegr}.

A possible explanation for not finding similar results to Kaufmann and Goodwin~\cite{Kaufmann2018} is that we use a more sophisticated measure of exposure than is common in the literature. Typically, exposure is measured as the portion of the group to which exposure is being measured within the unit of measurement (neighbourhood, postcode, municipality). Such a measure cannot take the variance in the spatial distribution of the group in question into account at all scales. The measure of exposure that we adapt from Brown and Enos~\cite{Brown2021} and use takes into account the distances between individuals. With this added distance information, the measure distinguishes the difference between a uniformly spread population and a highly segregated one.

\subsection{The role of unemployment}
The reported statistic in the CBS data for unemployment is in the form of the number of individuals receiving any of; an unemployment benefit, an assistance benefit, or a disability benefit. The data are often suppressed due to privacy concerns, so the population for which this statistic is reported is 1.3 million less than the entire population reported in the data. Due to their varied interpretations, we include them here as a separate analysis rather than in the main specification. We transform the raw number of benefit receivers into a share of the population in the `working' age (dividing by the sum of individuals aged 15-65).

We perform regressions with the share of people who receive benefits as a control and then as an interaction variable (interacting with exposure) and the coefficients of interest of these regression models (M15 and M16) are shown in Table~\ref{tab:benefit_reg}. We see that the share of benefits as a control variable is positively associated with right-wing voting. Furthermore, its inclusion does not weaken the strength of the relationship between exposure and right-wing voting but rather makes it stronger.

\begin{table}[htb]
\caption{Results of the multiple regression including the share of benefit recipients as control and as interaction variable. The remaining control variables (age group shares, address density, WOZ value and COROP fixed effects) are included but suppressed in the table due to focus. Standard errors in parentheses.}
\label{tab:benefit_reg}
\begin{center}
\begin{tabular}{l|ll}
\hline
 & {Benefits} & {Benefits interaction}\\
 & M15 & M16 \\
\hline
Exposure                      & -0.461*** & -0.264***     \\
                                & (0.032)   & (0.051)       \\[4pt]
% oad\_pop\_mean\_log\_scaled & -0.002     & -0.004           \\
%                             & (0.003)    & (0.003)          \\
% woz\_pop\_mean\_log\_scaled & -0.058***  & -0.059***        \\
%                             & (0.002)    & (0.002)          \\
% share\_15\_to\_25           & -0.131***  & -0.115***        \\
%                             & (0.039)    & (0.041)          \\
% share\_25\_to\_45           & 0.062**    & 0.065**          \\
%                             & (0.025)    & (0.025)          \\
% share\_45\_to\_65           & 0.489***   & 0.504***         \\
%                             & (0.055)    & (0.054)          \\
 $\texttt{S}_{WW}$             & 0.192***  & 0.339***      \\
                                & (0.051)   & (0.069)       \\[4pt]
Exposure:$\texttt{S}_{WW}$  &           & -1.412***     \\
                                &           & (0.274)       \\[10pt]
R-squared                       & 0.600     & 0.604         \\
R-squared Adj.                  & 0.594     & 0.598         \\
N                               & 2816      & 2816          \\
\hline
\multicolumn{3}{l}{*$p<.1$, **$p<.05$, ***$p<.01$}
\end{tabular}

% *$p<.1$, **$p<.05$, ***$p<.01$
\end{center}
\end{table}

The interaction between exposure and the share receiving benefits is strongly negative (in M16). This is the opposite of what one would expect in a situation where the local group feels threatened by immigrants in relation to competition for scarce resources. This coincides with a positive relationship between the share receiving benefits and right-wing voting (row $\texttt{S}_{WW}$ in M16). So, while the share of unemployed in a region increases right-wing voting in general, the presence of immigrants alleviates this rather than accentuates it.

\section{Voting behaviour and anti-immigrant sentiment}\label{app:liss_check}
In order to perform the desired analysis (comparing exposure to immigrants with anti-immigration sentiment) we need to link levels of anti-immigration sentiment to people of geographical areas. Some scholars perform the linking directly by associating survey respondents with their geographic location or by measuring their exposure to immigration directly (see, for instance, the network based exposure compared to anti-immigration sentiment studied by Kazmina \textit{et al.}~\cite{kazmina2024}). We take an indirect approach in order to bootstrap the responses of the LISS survey to almost the entire Dutch population.

The LISS Panel (see\url{https://www.dataarchive.lissdata.nl/study-units/view/22}) is a longitudinal survey conducted yearly since 2008. The survey is sent to more than 6000 participants each year, with a completion rate of 84\%. In particular, we use the Politics and Values section of this survey to verify a link between anti-immigration sentiment and voting behaviour in the Dutch context.

\subsection{Immigration questions}
The questions are posed in Likert-scale format asking ``What is your opinion on the following statements?'' with responses: \textit{1.\,fully disagree, 2.\,disagree, 3.\,neither agree nor disagree, 4.\,agree} and \textit{5.\,fully agree}.
The statements are:
\begin{itemize}
    \item 116. It is good if society consists of people from different cultures.
    \item 118. It should be made easier to obtain asylum in the Netherlands.

    \item 120. There are too many people of foreign origin or descent in the Netherlands.
    \item 123. It does not help a neighbourhood if many people of foreign origin or descent move in.
\end{itemize}
These questions are selected because they include a normative character alongside the descriptive nature. That is, they ask for more than an objective characterisation of the situation in the Netherlands. Question 121, asking for agreement with the statement ``{People of foreign origin or descent are not accepted in the Netherlands}'', can be interpreted as asking what the situation is like for immigrants rather than what the respondent feels the situation \textit{should be}. Similarly, question 122 asks about agreement with the statement ``{Some sectors of the economy can only continue to function because people of foreign origin or descent work there}'' yet does not include anything about whether respondents view that fact, if true, as good or bad.

\subsection{Responses to selected questions per voter group}
\begin{figure}[htb]
    \centering
        \begin{subfigure}{0.45\textwidth}
        \includegraphics[width=0.9\textwidth]{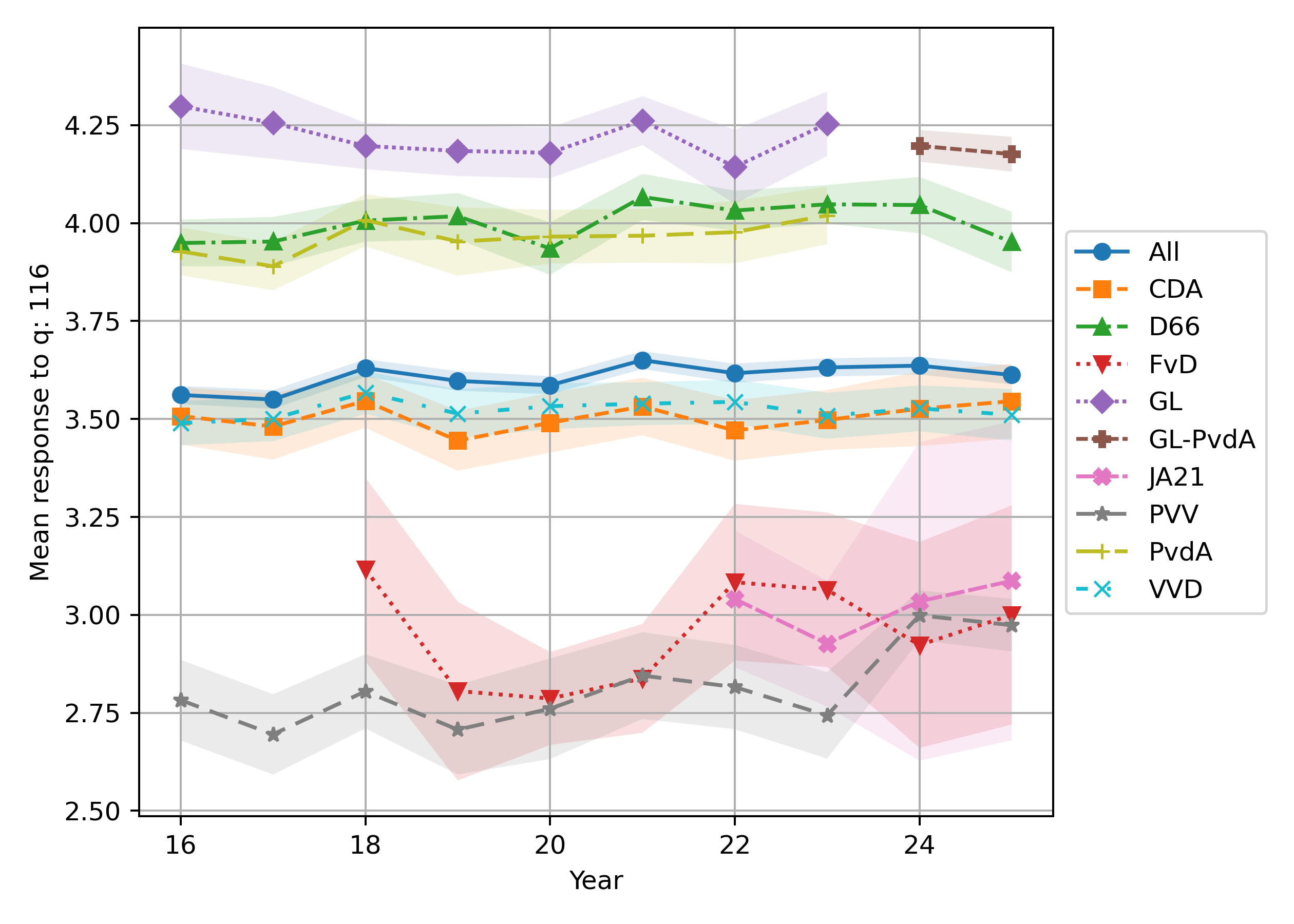}
        \caption{116. It is good if society consists of people from different cultures.}
    \end{subfigure}%
    \begin{subfigure}{0.45\textwidth}
        \includegraphics[width=0.9\textwidth]{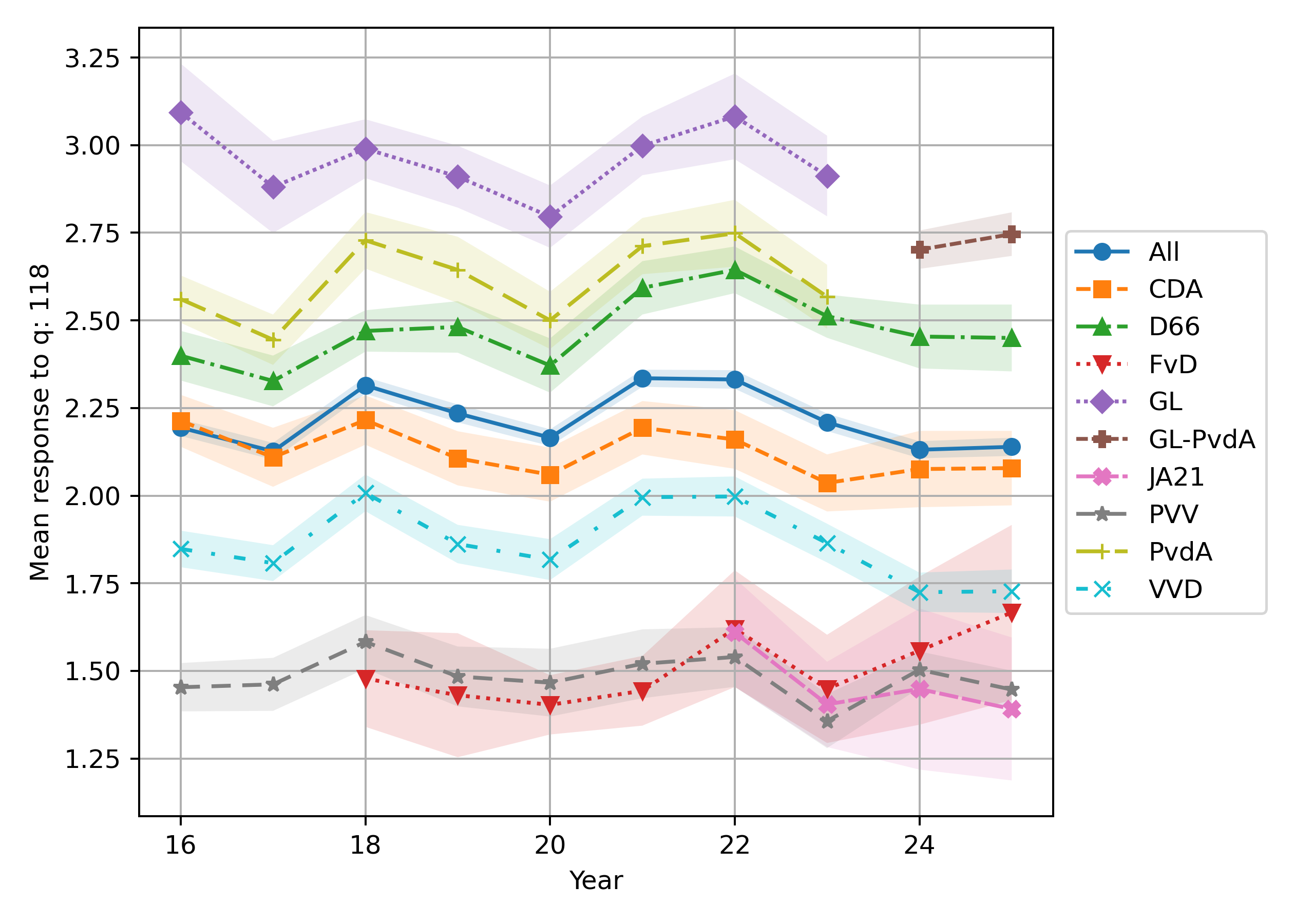}
        \caption{118. It should be made easier to obtain asylum in the Netherlands.}
    \end{subfigure}%
    \\
    \begin{subfigure}{0.45\textwidth}
        \includegraphics[width=0.9\textwidth]{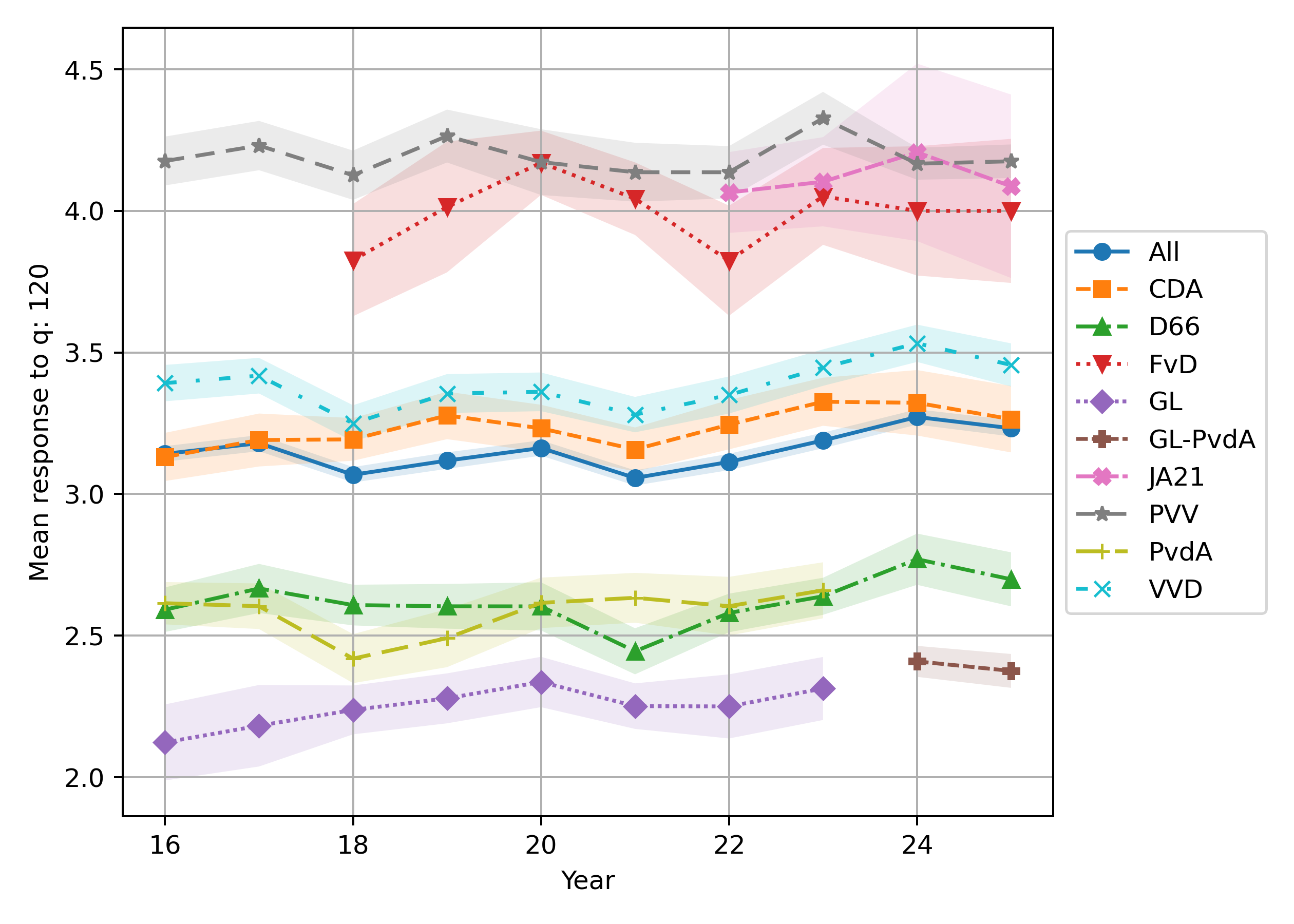}
        \caption{120. There are too many people of foreign origin or descent in the Netherlands.}
    \end{subfigure}%
    \begin{subfigure}{0.45\textwidth}
        \includegraphics[width=0.9\textwidth]{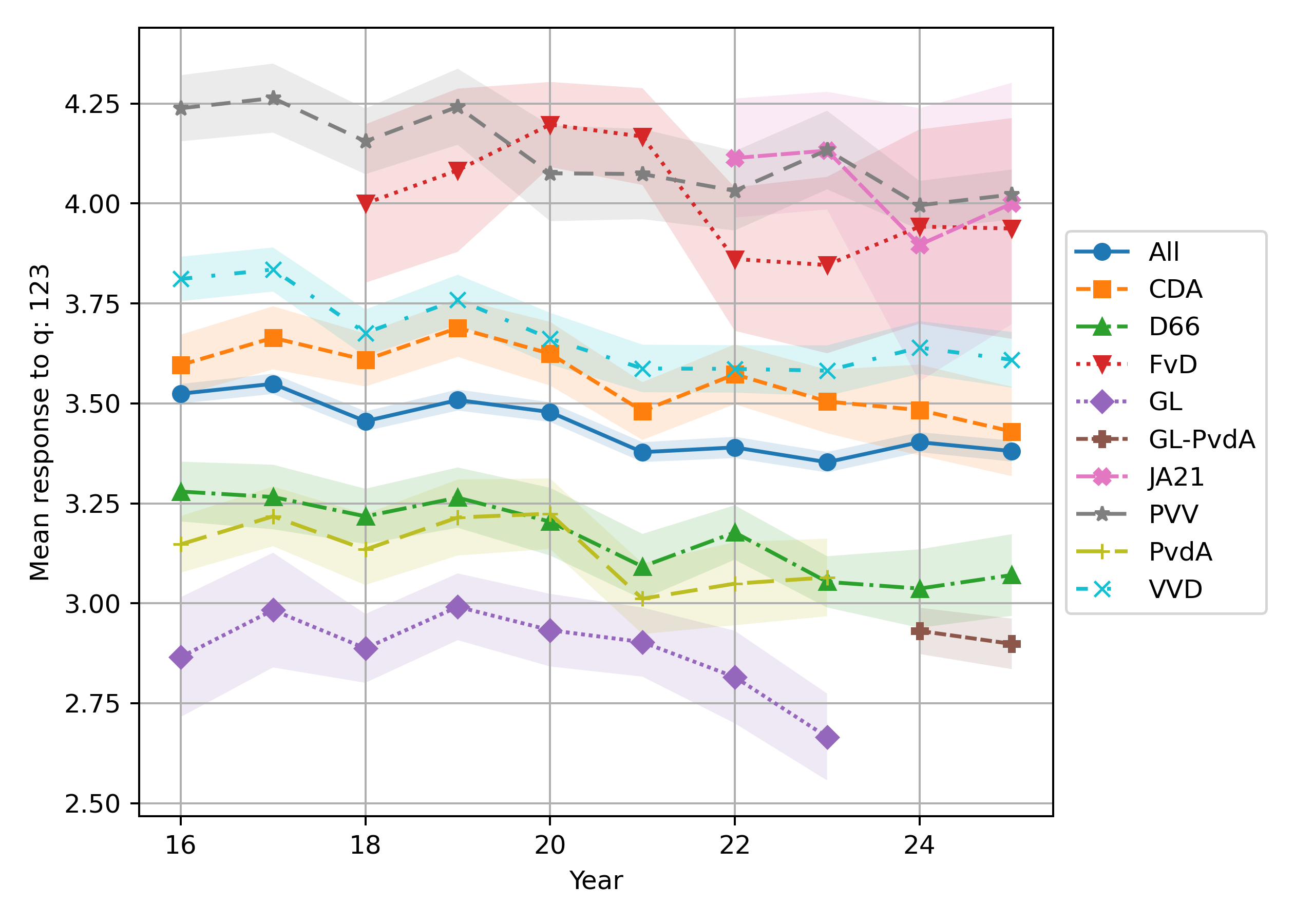}
        \caption{123. It does not help a neighbourhood if many people of foreign origin or descent move in.}
    \end{subfigure}%
    \caption{Mean value and 95\% confidence intervals for the responses to the questions related to immigration over time from 2016--2025.}
    \label{fig:time_questions}
\end{figure}

Figure~\ref{fig:time_questions} shows the average response (and 95\% confidence intervals) for responders according to which party they voted for in the most recent election (self-reported) for a the questions we take into consideration. A clear picture emerges in terms of party alignment and anti-immigration sentiment. Voters of PVV, FvD and JA21 consistently hold opinions which are more opposed to immigration than the more central voters for CDA, VVD. On the other hand, voters for GL, PvdA (and later GL-PvdA) and D66 consistently hold views that are more pro-immigration than the population average and the other clusters of voters. 

Particularly striking is panel (c) in Figure~\ref{fig:time_questions} plotting the mean agreement with the statement `There are too many people of foreign origin or descent in the Netherlands.' The responses to this question show a clear separation of the three clusters forming the major left, centre, and right parties. We also see that there is a slight increase in agreement among the left-wing parties and the total mean agreement. This points to an increase in the anti-immigration sentiment in the Netherlands from 2020, following the decrease from 2016 to 2020. 

Taking these results into account alongside the voting outcomes over the years 2021--2023 we feel confident taking the portion of votes awarded to $\mathcal{RW}:=\{PVV, FvD, JA21\}$ as a proxy for the level of anti-immigration sentiment.

\section*{Data and code availability statement}
All the data used are freely available to researchers, though are not owned by the authors. The data may be downloaded at:
\begin{itemize}
    \item The LISS panel data giving anti-immigrant sentiment and self-reported voting behaviour: \url{https://www.dataarchive.lissdata.nl/study-units/view/22}. A free account is required access the data.
    \item The population demographic data from the CBS (Statistics Netherlands) website: \url{https://www.cbs.nl/nl-nl/dossier/nederland-regionaal/geografische-data} under the heading `Kaart van 500 meter bij 500 meter met statistieken' for the 500 by 500 meter grid, and `Kaart van 100 meter bij 100 meter met statistieken' for the 100 by 100 metre grid.
    \item The election results are available from the Kiesraad: \url{https://www.verkiezingsuitslagen.nl/verkiezingen/detail/TK20231122}.
\end{itemize}

The code facilitating this analysis is available at \url{https://github.com/Benephfer/Exposure-and-anti-immigrant-sentiment-in-the-Netherlands}.

\section*{Competing interests}
The authors declare that there are no competing interests.

\section*{Ethics statement}

This paper performs secondary analysis of the survey data from the LISS panel, together with publicly available aggregate election results and small-area demographic data from Statistics Netherlands (CBS) and the Kiesraad (see the data and code availability statement above). The authors did not recruit, contact, compensate, or otherwise directly interact with human participants.

The LISS data were accessed under the conditions specified by the data provider. The study does not seek to identify individual respondents, and results are reported only in aggregate form. The research presents minimal risk to participants. 

\section*{Generative artificial intelligence statement}
The analyses in this paper were conceived and executed by the authors. Similarly the manuscript was written and prepared by the authors. Generative artificial intelligence (gpt-5.6-terra) was used to facilitate the writing and cleaning of code used in the analysis.

\section*{Acknowledgements}
The authors would like to acknowledge Jay Armas, Clelia de Mulatier, Wout Merbis, Frank Pijpers, Fernando Santos, Han van der Maas, Lourens Waldorp, Tuan Pham and Cees Diks for their feedback at the Polarisation, Segregation and Inequality research priority area meetings. The authors also wish to thank Bjarne Kosmeijer for his help in cleaning the election result data.

BM and TG are supported by the Dutch Institute for Emergent Phenomena (DIEP) cluster at the
University of Amsterdam under the Research Priority Area \emph{Emergent Phenomena in Society: Polarisation, Segregation, and Inequality}.”

\end{document}